\documentclass[sigconf]{acmart}

\usepackage{booktabs} 
\usepackage{multirow}
\usepackage{diagbox}

\setcopyright{acmcopyright}
\acmDOI{xx.xxx/xxx_x}

 \usepackage{amsmath}
 \usepackage{algorithmic}
 \usepackage{graphicx}
 \usepackage{algorithm}
 \usepackage{subcaption}

\begin{document}

\title{Energy Efficient Obfuscation of Side-Channel Leakage for Preventing Side-Channel Attacks}

\renewcommand{\shorttitle}{EE Obfuscation of SC Leakage for Preventing SCAs}


\author{Shan~Jin, Minghua Xu, and~Yiwei Cai}
\email{shajin@visa.com, mixu@visa.com, yicai@visa.com}
\affiliation{
  \institution{Visa Research}
  \city{Austin}
  \state{Texas}
  \country{USA}
  \postcode{78759}
}

\begin{abstract}
How to efficiently prevent side-channel attacks~(SCAs) on cryptographic implementations and devices has become an important problem in recent years. One of the widely used countermeasures to combat power consumption based SCAs is to inject indiscriminate random noise sequences into the raw leakage traces. However, this method leads to significant increases in the energy consumption which is unaffordable cost for battery powered devices, and ways must be found to reduce the amount of energy in noise generation while keeping the side-channel invisible. In this paper, we propose a practical approach of energy efficient noise generation to prevent SCAs. We first take advantage of sparsity of the information in the leakage traces, and prove the existence of energy efficient noise generation that is optimized in the side channel protection under a given energy consumption budget, and also provide the optimal solution. 
Compared to the previous approach that also focuses on the energy efficiency, our solution is applicable to all general categories of compression methods. 
Furthermore, we also propose a practical noise generator design by aggregating the noise generation patterns produced by compression methods from different categories. As a result, the protection method presented in this paper is practically more applicable than previous one. 
The experimental results also validate the effectiveness of our proposed scheme.

\end{abstract}

\ccsdesc[500]{Security and privacy~Tamper-proof and tamper-resistant designs}
\ccsdesc[500]{Security and privacy~Embedded systems security}

\keywords{
Side-Channel Attacks, Artificial Noise, Energy Efficiency, Channel Capacity}
\maketitle

\section{Introduction}
\label{sec:introduction}

\subsection{Background}
Side-channel attacks, which exploit the leaked information from physical traces such as power consumption to extract secret information~(keys, password, etc.) stored on the target device, have shown to be successful in attacking cryptographic operations and compromising the security of computer devices over the years. The effectiveness of side-channel attacks is based on the fact that the physical information emitted from the computer are dependent on the internal state (or secret) stored on the devices. This dependency is often called \emph{leakage model} or \emph{leakage function}~\cite{brier2004correlation}. There is a variety of forms of physical leakages, such as electromagnetic emanation~\cite{gandolfi2001electromagnetic}, acoustic emanation~\cite{faruque2016acoustic}, power consumption~\cite{kocher1999differential}, or others. Internet of Things~(IoT) and embedded devices, which are deployed in less protected location and are easily accessible physically, are more vulnerable to these forms of attacks. Typically, power consumption analysis is a widely studied method for SCAs. By analyzing the power consumed by a particular device during a computation, attackers attempt to infer the secret inside the device. 
In this paper, we will focus on power consumption based SCAs.


Fundamentally, side-channel attacks can be distinguished as non-profiled and profiled attacks. In~\emph{non-profiled} side-channel attacks, the attackers leverage an \emph{a-priori} assumption about the leakage model for extracting the secret. In power consumption based side-channel attacks, there are two widely used models in non-profiled attacks:~The \emph{Hamming Weight}~(HW) model~\cite{brier2004correlation} and the \emph{Hamming Distance}~(HD) model~\cite{doget2011univariate}. 
In \emph{profiled} side-channel attacks, the attacker first collects the leakage traces from the identical or from a similar device, and then learns the leakage model based on the collected data. This renders profiled side-channel attacks to be data-driven approaches. There are two commonly used techniques in profiled attacks, which are the \emph{Template Attack}~\cite{chari2003template,choudary2013efficient} and the \emph{Stochastic Model}~\cite{schindler2005stochastic}. Recently, the studies on Deep Learning based profiled attacks have also been made, such as~\cite{das2019cross,ryad2020deep}.   


\subsection{Related Work}
The effectiveness of SCAs is based on the fact that the physical leakage traces are correlated to the secret used in computation. Hence, breaking this correlation is an intuitive approach to counter SCAs. There are two main methods that have been discussed in the literature:~Masking~\cite{prouff2013masking}, and random noise generation~\cite{protect2000shamir}. 

Masking acts directly at the algorithmic level of the cryptographic implementation. The principle of this technique is to randomly split the secret into {multiple}~(for example $d+1$) \emph{shares}, where any $d$ shares can not reveal the sensitive information. In~\cite{prouff2013masking}, it was proved that given the leakage on each share, the bias of the secret decreases exponentially with the order of $d$. 
Although masking is shown to be an effective countermeasure to against SCAs, it requires a complete reconstruction of the cryptographic algorithm, such as the masked S-Box in AES Masking~\cite{thomas2016masking}, which is costly to implement, and impeding the direct usage of standard IP blocks. 

Different from masking, random noise injection based methods do not alter the core implementation of the cryptographic algorithm~\cite{protect2000shamir}. By superposing the random noises on the raw leakage traces, the Signal to Noise ratio~(SNR) of the received signal is decreased, which eventually hides the leakage information of the secret-dependent operations inside the system. However, as shown in~\cite{das2017attenuated}, this technique incurs a huge overhead~(nearly 4 times the AES current consumption) in order to obtain a high resistance to CPA~(Correlation Power Analysis)~\cite{brier2004correlation} based attacks. 
This motivates us for exploring more energy efficient noise generation approach to prevent side-channel attacks.


Besides, recent work in~\cite{fang2022scas} proposed a machine learning based approach to compensate the small signal power, for reducing the power overhead. However, this approach relied on applying the supervised machine learning, which brings a new issue of the cost in the implementation, especially to the small-sized IoT devices.   

Recently, the work in~\cite{Jin2022} also proposed an energy efficient noise generation scheme to prevent side-channel attacks. However, the proposed scheme only relies on the sample selection based compression methods~\cite{chari2003template} to extract the useful information from the leakage traces, which is not directly applicable to other linear compression cases such as PCA or LDA~\cite{choudary2013efficient,Archambeau2006subspace,standaert2008using,Zhou2017ANU}. Moreover, the designed noise generation scheme is based on a particular pre-defined sampling method at a time, which further reduces the applicability to protect the real world systems. In addition, for selecting the most suitable sample selection method, the cross-validation based strategy proposed in~\cite{Jin2022} is also costly to compute. Considering above limitations, it's expected to design a more practical energy efficient noise generation mechanism to combat side-channel attacks. 

\subsection{Summary and Contributions} 
In this paper, we present a practical approach to generate energy efficient noise sequence for combating against power based side-channel attacks. First, by modeling the leakage side-channel as a communication model, the mutual information between the secret and the side-channel leakage can be measured by the side-channel's channel capacity. 
Similarity to the case of secret communication, where the transmitter generates the artificial noise to degrade the channel capacity of the eavesdropper's channel, our goal is to generate artificial noise to degrade the side-channel's channel capacity for preventing the side-channel attacks.



Compared to previous random noise injection based methods, our method focuses on optimizing maximizing side-channel protection while minimizing the energy consumption on the system. Our method is based on the fact that information related to secrets is sparsely distributed among the leakage traces, and that for any budget of energy in noise generation, our method can derive a noise scheme that minimize the side channel's channel capacity, and our method achieves the~\emph{optimal energy efficiency}~(EE) for protecting the system from side-channel attacks. 


Furthermore, compared to the previous work in~\cite{Jin2022} which only relies on the sample selection based compression methods, our noise generation scheme can work for all general categories of compression methods. Besides, the assumption of a pre-defined sampling method in~\cite{Jin2022} is far less applicable as attackers can take different compression methods during attack. To address this issue, we also design a more practical noise generation mechanism by aggregating the optimal noise generation patterns conducted by a variety of compression methods from different categories. Without making any prior assumption on attacker's attack strategies, our design can achieve both strong security and applicability. 


\subsection{Paper Organization}
The remaining part of this paper is organized as follows. Section~\ref{sec:system-model} introduces the system model. The data compression is discussed in Section~\ref{sec:data-compression}. The optimal energy efficient noise generation is introduced in Section~\ref{sec:energy-efficient}. A practical design of the noise generator is presented in~\ref{sec:practical-design}. The further analysis is given in Section~\ref{sec:analysis}. 
Section~\ref{sec:experimental-results} presents the experimental results. The paper is concluded in Section~\ref{sec:conclusion}.

\section{System Model}
\label{sec:system-model}

\subsection{Modeling Side Channels as Communication Channels}
\label{sec:communication-channel}
A variety of previous work has focused on the approaches of modeling side-channel attacks model through communication theorems. In~\cite{standaert2009unified}, a general uniform framework is proposed to integrate SCAs and classic communication theory. 
In~\cite{heuser2014good}, by treating the side channel as a noisy channel in communication theory, the authors designed a number of optimal side-channel distinguishers for different scenarios. In~\cite{Jin2018}, the authors model the side-channel as a fading channel, which translates the profiling problem in side-channel attacks as a channel estimation problem in communication systems. In~\cite{eloi2019best}, by modeling side-channel as a communication channel, the mathematical link between the the probability of success of the attack and the minimum number of queries is derived. 

\begin{figure}
\centering
\includegraphics[width = 0.48 \textwidth]{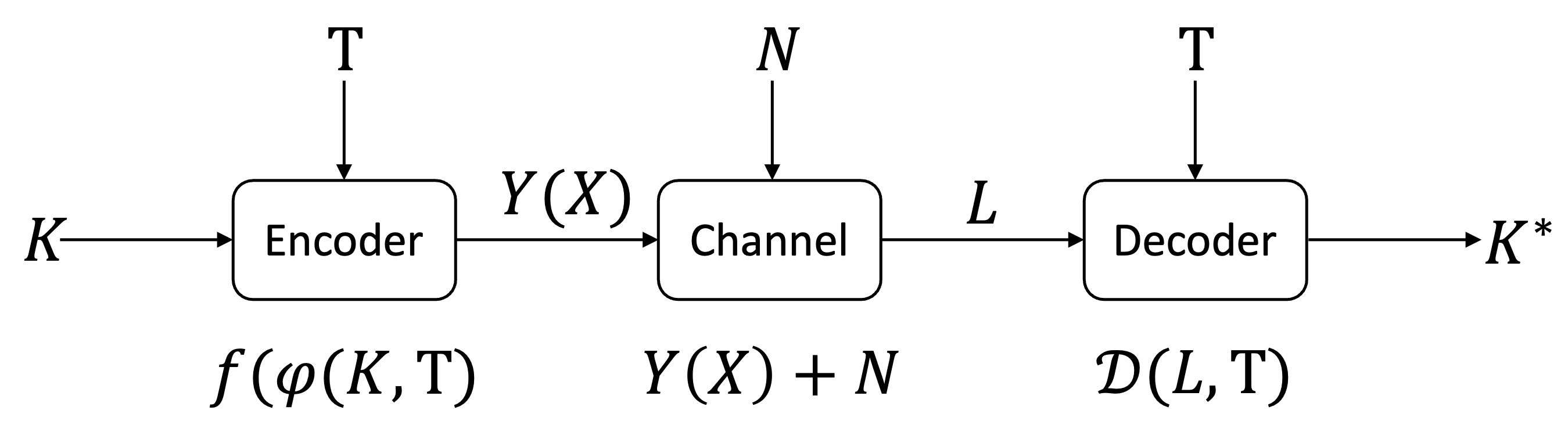} 
\caption{Side Channel Modeled as A Communication Channel}\label{fig:communication_channel}
\end{figure}

Generally, through the view of communication theory, as depicted in Fig.~\ref{fig:communication_channel}, the side-channel model can be modeled as:
\begin{align}
L=Y(X)+N~~,\label{eq1}
\end{align}
in where the components in the model are given as follows: 
\begin{itemize}
\item $K \in \mathbb{F}^{B}_2$ is the targeted secret key which is $B$ bits long~(for example $B=8$ for AES where attacker tries to recover each byte in a divide and conquer strategy). $\text{T} \in \mathbb{F}^{B}_2$ denotes an input plaintext or a received ciphertext with the size of $B$ bits. Typically, during the execution of a given cryptographic algorithm, a sequence of $I_q$ text bytes is inputted, which is represented by $\textbf{T}=(\text{T}_1,\dots,\text{T}_{I_q} )$.   
\item The \emph{encoder} is represented by any function of $f(\phi(K, \text{T}))$, which consists of two parts:~$\phi(\cdot)$ denotes a cryptographic operation which is defined by the algorithm's specification~(for example if the operation is the Sbox in AES:~$S(\cdot)$, we have $\phi(K, \text{T}) = S(K \oplus \text{T})$). We call $X$ the \emph{sensitive variable}~(or \emph{internal state}) where $X=\phi(K, \text{T})$;~$f(\cdot)\colon\mathbb{F}^{B}_2 \rightarrow \mathbb{R}^{m}$ is the leakage function, which is the deterministic part and is the mapping between the $X$ and a leakage trace $Y(X)$ which contains $m$ time samples:~$Y(X)=f(X)=f(\phi(K, \text{T})$. 
\item The \emph{side channel} is in fact an additive noise channel as modeled in~Eq.~\eqref{eq1}, where $N \in \mathbb{R}^{m}$ is the random part and typically is modeled as an i.i.d Gaussian noise vector with distribution $\mathcal{N}(0, \Sigma_{{N}})$ where $\Sigma_N$ is a diagonal matrix with identical diagonal value $\sigma^2_N$.
\item The \emph{decoder} is in fact a \emph{distinguisher} $\mathcal{D}$ to guess the target key:~$K^*=\mathcal{D}(L, \text{T})$. Usually, the attacker captures $I_q$ leakage measurements:~$\boldsymbol{L}=(L_1,\dots,L_{I_q} )$ which is obtained by inputting the plaintext vector $\textbf{T}$ to the implementation, to perform an efficient attack. This results in $K^*=\mathcal{D}(\boldsymbol{L}, \textbf{T})$.  
\end{itemize}


From information theory, \emph{mutual information}~($MI$)~\cite{leakage2018thomas,standaert2009unified} is commonly used to measure the amount of information related to secret $X$ that is contained in the leakage trace $L$. Since $f(\cdot)$ is a deterministic function of $X$, as derived in~\cite{eloi2019best}, we have
\begin{align}
& MI(L;X| \text{T}) = MI(L;Y(X)| \text{T})=\mathbb{E}[MI(L;Y(X)| \text{T}=t)] \nonumber \\
& \qquad \qquad \qquad \qquad \qquad \quad \ \ \ =H(Y(X))-H(Y(X)|L)~~,
\end{align}\label{eq2}
where $\mathbb{E}[\cdot]$ is the expectation operator, $H(Y(X))$ is the entropy of $Y(X)$ and $H(Y(X)|L)$ is the conditional entropy of $Y(X)$ based on $L$. The maximum mutual information leaked through a side-channel is typically called \emph{channel capacity}~\cite{elements2006cover}, which is expressed as:
\begin{align}
C=\max MI(L;Y(X)| \text{T})=\max MI(L;Y(X))~~.
\end{align}
Note here $C$ is measured by single trace. When $I_q$ traces are recorded, the channel capacity becomes $I_q\cdot{C}$~\cite{Cherisey2019information,cheng22masked}.

We note a few differences between SCAs model and communication model:~In SCAs, the leakage trace is measured~(received) by the attacker. However, rather than acting as the receiver, now the attacker in fact plays the role of the eavesdropper as in the communication model~\cite{goel2008guarantee}. Hence, the attackers can not influence the internal state $X$ in the target implementation~(transmitter) through the communications between the transmitter and the attackers as in the communication model~(this is obvious as now $X$ is the unknown target secret), and therefore can not optimize the channel capacity of the side-channel. Previous work in~\cite{leakage2018thomas} also focused on the same problem, where the authors construct the side-channel's channel capacity by modeling the side-channel as MIMO~(multi-input multi-output) channels~\cite{gold2003capacity}. However, the modeling of the side-channel does not have to be limited to the linear model. On the contrary, A variety of non-linear models have also been proposed and fully studied, such as convolutional neural networks~\cite{ryad2020deep} and the quadratic model~\cite{doget2011univariate}. In this paper, similarity to the work in~\cite{heuser2014good,Cherisey2019information,eloi2019best}, we use a general model to characterize the side-channel. We treat the side-channel as an additive \emph{noisy channel}, as shown in Eq.~\eqref{eq1}. Hence the channel capacity of the side-channel is expressed by~\cite{elements2006cover}: 
\begin{align}
C=\frac{1}{2}\log\left(1+ \text{SNR}\right)=\frac{1}{2}\log\left(1+\frac{\|Y(X)\|^2_2}{\mathbb{E}[\|N\|^2_2]}\right)~~,\label{eq3}
\end{align}
where $\|\cdot\|_2$ is the $\ell_2$ norm. 
We also note that Eq.~\eqref{eq3} is applicable to all forms of leakage models, despite HW or HD models, the linear models, or any non-linear models, given the recorded side-channel leakage $Y(X)$.

\subsection{Using Random Noise to Counter SCAs}
\label{sec:random_noise}
As shown in Eq.~\eqref{eq3}, the channel capacity is decided by the SNR of the received leakage signals. Hence, mitigating the {effectiveness} of the SCAs is equivalent to reducing the SNR of the leakage signals. In communication system, one effective way to reduce the SNR is to inject random noise into the transmitted signal, which therefore decreases the SNR of the signal received by the eavesdropper, and leads to the degrading of the eavesdropper's channel capacity. This technique is generally referred as \emph{jamming}~\cite{sha2005jam}, and has also been proven to be effective in combating SCAs~\cite{protect2000shamir,generic2011tim}.



\begin{figure}
\centering
\includegraphics[width = 0.48 \textwidth]{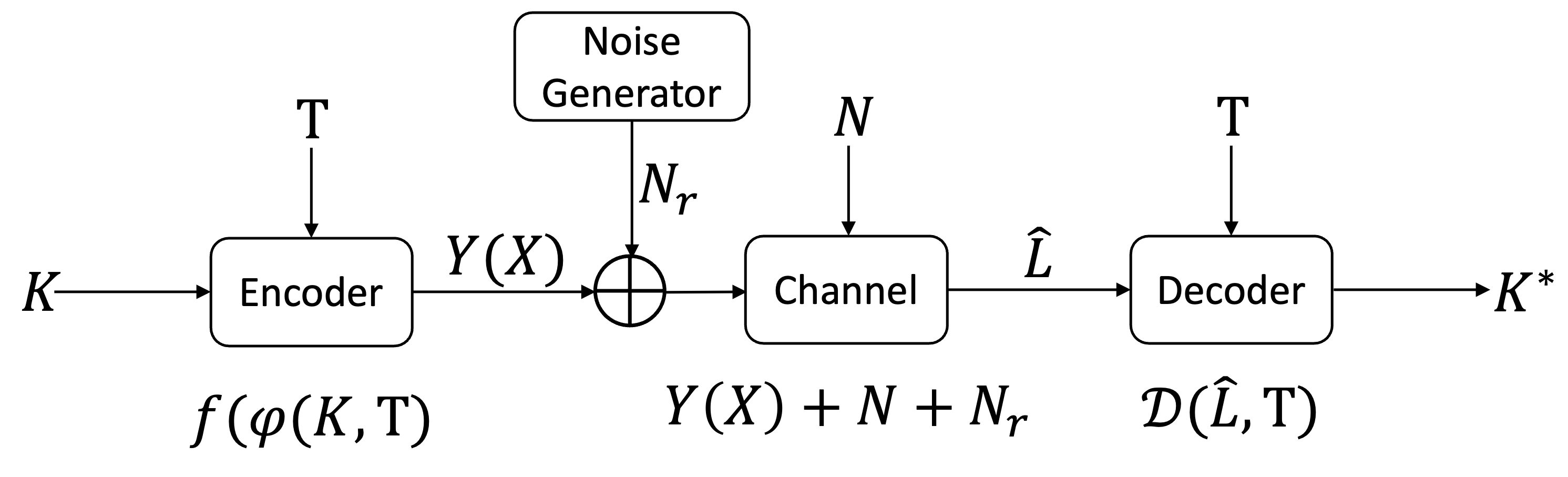} 
\caption{Side Channel Model with Random Noise Injection}\label{fig:noise_generator}
\end{figure}

Fig.~\ref{fig:noise_generator} shows the side-channel model after applying the jamming. 
The random noise sequence is independently and randomly generated from the \emph{noise generator}, which is then injected into the raw leakage trace $Y(X)$. 
As a result, the raw leakage trace and the random noise sequence are mixed together, which results in a ``new'' leakage trace that measured by attacker. Generally, the leakage traces superposed by the random noise can be expressed as:
\begin{align}
\widehat{L}=Y(X)+N+N_r~~,
\end{align}
where $N_r$ is the i.i.d Gaussian noise vector with a distribution of $\mathcal{N}(0, \Sigma_{{r}})$. Now the capacity of the side-channel becomes:
\begin{align}
C=\frac{1}{2}\log\left(1+\frac{\|Y(X)\|^2_2}{\mathbb{E}[\|N\|^2_2] + \mathbb{E}[\|N_r\|^2_2]}\right)~~.\label{eq4}
\end{align}
Compared to Eq.~\eqref{eq3}, the additional term $\mathbb{E}[\|N_r\|^2_2]$ decreases the SNR in Eq.~\eqref{eq4} as well as the channel capacity $C$. As discussed in~\cite{eloi2019best,Cherisey2019information}, given a fixed $P_s$ which is the probability of success of an attack:~$\mathbb{P}(K^*=K)$, the required minimum number of quires (number of traces) $I_q$ is inversely related to the SNR:~$I^{\text{min}}_q \propto \frac{1}{\log(1+\text{SNR})}$. This means in order to achieve the same $P_s$, now attacker needs to record much more leakage traces, compared to the original SCAs. 


Although the random noise generation method clearly works and is easily to be implemented, previous work also shows that this method leads to significant energy consumption:~In~\cite{das2017attenuated}, the authors measured that in order to get a high enough resistance to side-channel attacks on AES, the current consumption is increased by a factor of four. This motivates us to design an energy efficient and also powerful noise generation scheme to counter SCAs.

\section{Data Compression Technologies}
\label{sec:data-compression}

Before introducing the design of our proposed scheme, we first discuss the compression of the leakage signal during the profiling and the attack phases. Typically, due to high sampling rates used, the recorded actual leakage trace $L_o$ has a very large number $m_o$ of samples where $m_o \gg m$~(note the recorded samples has already been composed of the same sized random part $N$~or even injected random noise $N_r$). However, it has been observed (such as in~\cite{chari2003template,Archambeau2006subspace,choudary2013efficient}) that leakage traces are \emph{sparse}, i.e. only very few leakage samples in the raw leakage traces contain useful information that is related to the internal state. This sparsity is caused by the fact that only a few logical components in the cryptographic system are depending on the secret in the predefined Input/Output~(I/O) activities. 
Thus, in order to perform an effective attack, typically attacker has to {\em compress} the leakage traces in order extract the useful leakage information for attacking.

Based on how the samples are processed, typically the compression methods can be divided into two categories:
\begin{itemize}
 \item \emph{Sample Selection}:~By selecting a subset from all recorded leakage samples, the number of samples is reduced. Samples are selected by using a variety of methods including Difference of Means (DoM)~\cite{chari2003template}~(which includes 1ppc, 3ppc, 20ppc and allap), Sum of Squared Pairwise T-Differences (SOST)~\cite{gierlichs2006templates}, or Signal-to-Noise Ratio~(SNR)~\cite{mangard2008power}.
  \item \emph{Linear Combination}:~Using a linear transformation for projecting the raw leakage samples onto a low-dimensional subspace. Typically, this is done by using Principal Component Analysis~(PCA) and Linear Discriminant Analysis~(LDA) \cite{choudary2013efficient,Archambeau2006subspace,standaert2008using,Zhou2017ANU}.
\end{itemize}
Mathematically, independent from which methods is applied, the process of compression can be expressed as:~$P{L_o}$,
where $P \in \mathbb{R}^{m_c \times m_o}$ with $ m_c \leq m$ is the {\em compression matrix}. The structure of the matrix $P$ corresponds to the category of compression methods:

\paragraph{Sample Selection}~In this scenario, the compression matrix is also called the {\em sampling matrix}. Each row in $P$ has exact one "1", and the index of the "1" in each corresponding column of $P$ stands for which sample in the recorded leakage signal is to be picked. (We can think of the sampling matrix as a permutation matrix with additional zero columns.). 

\paragraph{Linear Combination}~Any of the linear combination approaches uses a projection matrix, call it $\boldsymbol{U}$, to transform the raw leakage sample vector into a lower-dimensional vector. In PCA, $\boldsymbol{U}$ is composed of the first $m_c$ eigenvectors after applying the singular value decomposition~(SVD) on the \emph{sample between groups
matrix} $B$ by selecting the top $m_c$ components;~while in LDA, $\boldsymbol{U}$ is the first $m_c$ eigenvectors of $S^{-1}_{\text{pooled}}B$ where $S_{\text{pooled}}$ is the the \emph{common covariance of all groups}~\cite{choudary2013efficient,Archambeau2006subspace}~(same as~\cite{choudary2013efficient}, here we also assume $S_{\text{pooled}}$ is appropriately scaled hence the covariance between discriminants is the identity matrix). In the following, we treat $P=U$ for all the linear combination based methods. 

However, no matter which compression method is chosen, the following observations for matrix $P$ are always held:
\begin{enumerate}
     \item \textbf{Right Inverse}:~We have $P \cdot P^{T} = I_{m_c}$ for all types of $P$, where $I_{m_c} \in \mathbb{R}^{m_c \times m_c}$ is an identity matrix.
     \item \textbf{Idempotent}:~Let $\hat{P} = P^{T}P$, we have $\hat{P}\cdot\hat{P} = \hat{P}$ for all types of $P$, where $\hat{P} \in \mathbb{R}^{m_o \times m_o}$ is a symmetric matrix. 
\end{enumerate}
These two observations will be used in the following to derive the energy efficient scheme.



\section{Optimal Energy Efficient Artificial Noise Design}
\label{sec:energy-efficient}

\subsection{System Model of Artificial Noise Generation}
\label{sec:artificila-noise_model}
In our designed model, the artificial noise sequence is also generated from the noise generator and injected into the raw leakage trace~(same as in Fig.~\ref{fig:noise_generator}).  
The generated noise is denoted as $\widehat{N_a}$, which is defined by:
\begin{align}
\widehat{N_a}=\rho F N_a~~,\label{eq10}
\end{align}
where $\rho$ is the gain factor, and $N_a \in \mathbb{R}^{m_o}$ is an i.i.d Gaussian distributed noise vector with the distribution $\mathcal{N}(0, \Sigma_{{a}})$ where $\Sigma_a$ is a diagonal matrix with identical value $\sigma^2_a$ on the diagonal. 

In Eq.~\eqref{eq10}, $N_a$ is the noise source of the \emph{noise generator}, and $\rho$ is the \emph{gain} on the amplitude of the generated impulsive noise. As we mentioned before, random noise generation is not energy efficient as it generates noises to corrupt all leakage samples. As introduced in Section~\ref{sec:data-compression}, typically only a subset of leakage samples contain secret-dependent useful information. Hence, our designed noise sequence which aims at only corrupting those useful
sample is in fact a \emph{impulsive noise sequence}. Moreover, the artificial noise generator can directly apply the majority modules in the traditional \emph{impulsive noise generator}~\cite{mann2002impulse}, but only makes the change on the modeling of the inter-arrival times of the noise sequence:~Different to the traditional impulsive noise generation, where the inter-arrival times of the impulsive noise is modeled as a \emph{Markov renewal process}~(MRP)~\cite{mann2002impulse,levery2002statistical} and hence the inter-arrival time between two consecutive impulses is random~(controlled by a Markov state-transition probability matrix), the inter-arrival times of our designed impulse noise is optimized given pre-defined energy overhead budget. 
More specifically, by explicitly control the generation of the noise, the impulsive noise sequence can superpose with the targeted information pattern contained in the leakage trace optimally, hence maximize the system security under SCAs.  

Typically, the generation pattern could be represented by a state-transition probability matrix, and is stored and triggered in the control module in the noise generator. The transition matrix is expressed by $G=[\beta_{i,j}]$ with $1 \leq i, j \leq n$,
where $n$ is the total number of states in the impulsive noise generation, $\beta_{i,j}$ is the transition probability from state $s_i$ to $s_j$, and the state $s_i$ represents where the $i$-th impulse event should be occurred in the sequence given the first $i-1$ impulses is optimized. Hence $n$ equals the maximum number of the impulses that could be generated in a duration. As we mentioned that the generation of the impulsive noise is optimized in our design, we have $\beta_{i,j} = 1$ only if $s_i$ matches with the first $i$ impulses in $s_j$~($s_i = s_{j-1}$), and $\beta_{i,j} = 0$ for the rest of the cases. Here $G$ is in fact a \emph{sparse matrix} since majority of the elements in $G$ is 0. 

Actually, the structure of the transition matrix can be mapped to a process of impulsive noise samples selection. This selection of the generation of the impulsive noises~(which can also be viewed as a set of instructions to trigger the impulsive generation or not) is represented by the diagonal matrix $F \in \mathbb{R}^{m_o \times m_o}$, which injects the impulsive noises into a subset of samples of the raw leakage trace. Values on the diagonal can be "0" or "1", and the positions of the``1''s in the diagonal decide where the noise samples will be generated and injected into the leakage trace, i.e., for which samples in the leakage trace we inject the impulsive noises. 

In contrast to the indiscriminate random noise generation approach that described in Section~\ref{sec:system-model}, the artificial noise generation model can exploit the fact of the sparsity in leakage trace as shown in Section~\ref{sec:data-compression}, and only generate a small number of noise samples at the relevant timing to inject into the raw leakage trace. This brings the basis for us to design more energy-efficient scheme to generate noise sequence. As we can notice, the key point work is to design an appropriated $F$ to achieve the optimal energy-efficiency, which will be discussed in the following. 




\subsection{Optimization Problem in Artificial Noise Generation}

Assuming now the recorded leakage trace is perturbed by an artificial noise sequence 
$\widehat{N_a}$, and a data compression method $P$ is selected, then after the data compression, the recorded leakage trace can be expressed by:
\begin{align}
  P\widehat{L_o}=Y_{P}(X)+N_{P}+P\rho F {N_a}~~,
\end{align}
where $Y_P(X)$ is the deterministic part and $N_P=PN$ is the random part noise, after compression. Then the channel capacity becomes:
\begin{align}
C=\frac{1}{2}\log\left(1+\frac{\|Y_P(X)\|^2_2}{\rho^2\mathbb{E}[\|PF{N_a}\|^2_2] + \mathbb{E}[\|{N}_P\|^2_2]}\right)~~,\label{eq11}
\end{align}
From the system model in Fig.~\ref{fig:noise_generator}, the operations in the noise generator are independent to the encoder in where the internal state is stored. Hence, the generation of the designed artificial noise sequence is based on the entire internal state space in the target implementation, and not specifically to any single internal state $X$, since we like to maximize the utilization of the protection for a wide spectrum of attacking techniques. 
We therefore take the expectation on $\|Y_P(X)\|_2$ over all states, which results in a constant ${E}_X=\mathbb{E}[\|Y_P(X)\|^2_2]$. We can therefore re-define the target channel capacity of the side-channel as
\begin{align}
 C_s=\frac{1}{2}\log\left(1+ \frac{E_X}{\rho^2\mathbb{E}[\|PF{N_a}\|^2_2] + \mathbb{E}[\|{N}_P\|^2_2]}\right) ~~.\label{eq11_1}
\end{align}

As noted in ~\cite{goel2008guarantee}, in order to guarantee the confidentiality of a secret communication, the noise injected into the channel must be designed to maximally degrade the eavesdropper's channel. More specifically, the objective in~\cite{goel2008guarantee} is to generate an artificial noise that maximizes the lower bound on the {\em secrecy capacity}, which is defined as the difference in mutual information between the transmitter-receiver pair and the transmitter-eavesdropper pair:~$\widetilde{C_{sec}} = MI_{TR} - MI_{TE}$
where $\widetilde{C_{sec}}$ is the lower bound on the secrecy capacity, $MI_{TR}$ and $MI_{TE}$ are the mutual information on the transmitter-receiver and the transmitter-eavesdropper pairs, respectively.

However, there is small difference when applying above condition to SCAs models. In SCAs, there is no receiver as in the communication model, and we therefore don't have to consider the transmitter-receiver pair. As a result, we can treat the mutual information in the transmitter-receiver pair as a constant value~(for example let $MI_{TR}=0$). Hence, from the view of secret communication, \emph{maximizing} the secrecy capacity in side-channel attacks is equivalent to \emph{minimizing} the channel capacity in the transmitter-attacker~(eavesdropper) pair, which happens to be correspond to the side channel in our problems. Therefore, our objective is to construct the noise sequence $\widehat{N_a}$ that can minimize the channel capacity $C_s$ of the side channel constraint by a threshold $E_A$ that caps the energy consumption for the noise generation: 
\begin{align}
\min_{\widehat{N_a}} C_s \qquad \text{s.t.} \ \mathbb{E}[\|\widehat{N_a}\|_2^2] \leq E_A~~.\label{eq12}
\end{align}

Noticed in Eq.~\eqref{eq10}, $F$ is used to control the generation of the artificial impulsive noise. More specifically, the rank of $F$ decides how many impulsive noise samples will be generated, which also decides the total energy to be spent in the noise generation. Hence the objective problem in Eq.~\eqref{eq12} can be translated into:
\begin{align}
\min_{F} C_s, \qquad \text{s.t.} \ \text{Rank}(F) \leq R_A~~,\label{eq13}
\end{align}
where $\text{Rank}(\cdot)$ counts the rank of a given matrix, and $R_A$ is the constraint on the number of impulsive noises that can be generated by the noise generator:~$R_A=\frac{m_oE_A}{\rho^2\text{Tr}(\mathbb{E}[{N_a}{N_a}^T])}$, which is proportional to the amount of energy required to generate the impulsive noises. Note typically we take floor operation on $R_A$~($R_A=\lfloor R_A \rfloor$) for enforcing it to be an integer. 


\subsection{Solution with Optimal Energy Efficiency}
Since the independent noise $\text{E}[\|{N}_P\|^2_2]$ is fixed once $P$ is chosen, we only need to focus on the artificial noise part. From the Eq.~\eqref{eq11_1}, we can further obtain
\begin{align}
 \mathbb{E}[\|PF{N_a}\|^2_2] =\mathbb{E}[{N_a}^TF^TP^TPF{N_a}]=\mathbb{E}[{N_a}^TF^T\hat{P}F{N_a}]~~,
\end{align}
Letting $\Delta = F^T\hat{P}F$, based on the matrix trace operation, we have
 \begin{align}
\rho^2\mathbb{E}[\|PF{N}_a\|^2_2] =\rho^2\mathbb{E}[\text{Tr}({N_a}^T\Delta {N_a})]=\rho^2\text{Tr}(\mathbb{E}[\Delta {N_a}{N_a}^T])~~, \label{eq14}
 \end{align}
where $\text{Tr}(\cdot)$ is the trace operation. 
Furthermore, we have  
 \begin{align}
\rho^2\text{Tr}(\mathbb{E}[\Delta {N_a}{N_a}^T])= \rho^2\sigma^2_a\sum^{m_o}_{i=1}F_{ii}\hat{p}_{ii}~~, \label{eq14_2}
 \end{align}
where $\hat{p}_{ii}$ and $F_{ii}$ are the $i$-th element on $\hat{P}$'s and $F$'s diagonal, respectively. 
Due to the monotonicity of the logarithmic function, one can easily find that minimizing $C_s$ in Eq.~\eqref{eq13} is equivalent to maximizing the energy spent in the noise generation~(the Eq.~\eqref{eq14_2}). Further, the maximization of Eq.~\eqref{eq14_2} under the energy constraint is in fact a typical 0-1 integer programming problem~\cite{Nemhauser1988integer}. Therefore the problem in Eq.~\eqref{eq13} can be translated as:
 \begin{align}
\qquad \qquad \max_{F} \rho^2\sigma^2_a\hat{P}^{T}_{d}F_{d} \quad \text{s.t.} \ AF_{d} \leq R_A, \ F_{d} \in \{0, 1\}^{m_o}~~,\label{eq15}
 \end{align}
where $\hat{P}_d \in \mathbb{R}^{m_o}$ and $F_d$ are the column vectors of the main diagonal elements of $\hat{P}$ and $F$ respectively, and row vector $A=[1,\dots,1] \in \mathbb{Z}^{1 \times m_o}$. In~Eq.~\eqref{eq15}, it's clear that both $\rho^2$ and $\sigma^2_a$ are non-negative values. Since $\hat{P}$ is a symmetric idempotent matrix, its diagonal elements always satisfy $\hat{p}_{ii} \in [0, 1]$. Hence, the objective in Eq.~\eqref{eq15} is monotonically increasing with the increasing of $|\Omega_F|$ which is the cardinality of $\Omega_F$, where $\Omega_F$ is the set of the indices of "1"s in $F$'s diagonal. Furthermore, when $|\Omega_F|$ is fixed~($|\Omega_F|=R_A$), based on the branch and bound method~\cite{Nemhauser1988integer}, the maximum is decided by those most significant $\hat{p}_{ii}$ that are selected by $\Omega_F$. 

More specifically, let's first define $\Omega_{\hat{P}}$ as a set of the indices of $\hat{P}$'s diagonal elements, sorted in a decreasing order by $\hat{p}_{ii}$'s value. This means $\Omega_{\hat{P}}[1]$ is the index of the largest element in $\hat{P}$'s diagonal and $\Omega_{\hat{P}}[m_o]$ is the index of the smallest one. Moreover, $\Omega_{\hat{P}}[1:k]$ represents a subset of $\Omega_{\hat{P}}$ where only the first $k$ elements in $\Omega_{\hat{P}}$ are included. However, to an arbitrary set $\Omega_{\hat{P}}$, we are in fact only interested in those non-zero diagonal elements. Therefore, we further define a mapping $\Psi_{\hat{P}}(k)$ that outputs a set with the largest number of indices that corresponding to the non-zero diagonal element values, given an input subset $\Omega_{\hat{P}}[1:k]$, as:
\begin{equation}
 \Psi_{\hat{P}}(k)=\left\{ 
  \begin{array}{ll}
       \Omega_{\hat{P}}[1:k], \quad  \text{if} \ \Omega_{\hat{P}}[k]>0~~, \\
      \Omega_{\hat{P}}[1:j], \quad   \text{if} \ \Omega_{\hat{P}}[j] >0  \ \& \ \Omega_{\hat{P}}[j+1]=0~~, \\ 
 \qquad \qquad \qquad \text{with} \ j<k~~.
 \end{array}
  \right.\label{eq16}
\end{equation}
Based on Eq.~\eqref{eq16}, we can obtain the optimum $F$ for the optimization problem in Eq.~\eqref{eq15}, which is
\begin{align}
F^*=\{ F \ | \ \Omega_F = \Psi_{\hat{P}}(R_A),\ |\Omega_F|\leq R_A  \}~~.\label{eq17}
\end{align}
Note when $P$ is a sampling matrix, $\hat{p}_{ii}$ is either 0 or 1. Hence the order of the indices of the "1"s in $\Omega_{\hat{P}}$ is randomly generated. Here one can easily find the main result in~\cite{Jin2022}~(Eq.~(24) in~\cite{Jin2022}) is in fact just a special case of Eq.~\eqref{eq17} in our work.

Eq.~\eqref{eq17} gives a solution to the design of an optimal energy-efficient artificial noise sequence under a given energy constraint $R_A$ and compression method $P$. However, in practice, normally there would be no prior information about which signal processing technique~(the $P$) that attackers will use. For example, the system may choose 20ppc to construct the artificial noise during the design phase, but the attacker may prefer to use 3ppc. We will discuss how to apply our theoretical result into practical usage in next section.






\section{A Practical Design of the Noise Generator}
\label{sec:practical-design}

Following the optimal solution derived in Sec.~\ref{sec:energy-efficient}, we propose an algorithm to construct a more practical energy-efficient noise generation mechanism for the system. Generally speaking, our algorithm is based on the philosophy of majority voting:~First, during the design phase, the system designer picks a set of candidate compression methods:~$\{P_i\}^{\zeta}_{i=1}$ with $\zeta$ is the total number of the selected compression methods. Based on Eq.~\eqref{eq17}, the corresponding optimums $\{\Omega_{F^*_i}\}^{\zeta}_{i=1}$ and $\{F^*_i\}^{\zeta}_{i=1}$ are therefore obtained, under a given energy constraint $R_A$. Then for each coordinate $i \in \{1,2,\dots,m_o\}$, if $i$ is labelled by the majority~($\geq 50\%$) sets of $\{\Omega_{F^*_i}\}^{\zeta}_{i=1}$, $i$ is selected into the final set $\Omega_{F_{o}}$. The final selection matrix $F^*_o$ that will be used in noise generation in practice is hence defined as
\begin{align}
F^*_o=\{ F_o \ | \ \Omega_{F_{o}} = \boldsymbol{\psi}, \ |\Omega_{F_{o}}|\leq R_A  \}~~. \label{eq171}
\end{align}
More details of our algorithm is illustrated in \textbf{Algorithm~1} and \textbf{Algorithm~2}. Once we obtain the final design $F^*_o$, we can easily translate $F^*_o$ to the state-transition matrix $G^*$ that introduced in Sec.~\ref{sec:artificila-noise_model}:~For $i = 1, 2, \dots, m_o$, if the $i$-th element in $F^*$'s diagonal vector $F^*_d$ has: ~$F^*_d[i]==1$, we set $s_i = F^*_o[1:i]$. Eventually, based on the states $s_i$~($i=1,\dots,m_o$), transition matrix $G$ is reconstructed.


{\renewcommand\baselinestretch{1}\selectfont
\begin{algorithm}
  \label{alg:1}
  \caption{Majority Voting based Coordinate Selection}
 \begin{algorithmic}
  \REQUIRE $\{\Omega_{F^*_i}\}^{\zeta}_{i=1}$
  \STATE  \textbf{Initialize}  \ $\ \boldsymbol{\psi}=\{1, 2, \ldots, m_o\}$
 \STATE \textbf{For} $j =1, 2,..., m_o$
  \STATE \quad $\textbf{If} \ \textbf{Majority\textunderscore Voting}(j, \  \{\Omega_{F^*_i}\}^{\zeta}_{i=1}) == 0$
  \STATE \qquad $\boldsymbol{\psi}=\boldsymbol{\psi} \backslash j;$
  \STATE $\textbf{end}$
\ENSURE $\boldsymbol{\psi}$
\end{algorithmic}
\end{algorithm}
\par}

{\renewcommand\baselinestretch{1}\selectfont
\begin{algorithm}
  \label{alg:1}
  \caption{\textbf{Majority\textunderscore Voting}}
 \begin{algorithmic}
  \REQUIRE $ j, \ \{\Omega_{F^*_{i}}\}_{i=1}^{\zeta}$
  \STATE  \textbf{Initialize}  \ $ t=0, \ \rho=0$
 \STATE \textbf{For} $l =1, 2,..., \zeta$
  \STATE \quad $\textbf{If} \ j \in \Omega_{F^*_{l}}$
  \STATE \qquad $t=t+1;$
  \STATE $\textbf{end}$
  \STATE $\rho = \text{Round}(t/\zeta);$
\ENSURE $\rho$
\end{algorithmic}
\end{algorithm}
\par}

\section{Analysis of the Noise Sequence Generation Scheme}
\label{sec:analysis}


\subsection{Energy Efficiency}
As we design noise generation in an energy efficient fashion, we need to be guided by a metric that informs us both of their {\em effectiveness} and the {\em efficiency}. The former measures in terms of the reduction of the successfulness of an attack, and the second measures in terms of the amount of energy that is consumed. We combine these two metrics together, and call it {\em energy efficiency}~(EE), and use it to measure the reduction of the attack success rate per unit of energy spent in noise generation, which is expressed by:
\begin{align}
{EE}=\frac{ 1 - {Successful\textunderscore Key\textunderscore Recovery} }{Noise\textunderscore Energy}~~,\label{eq18}
\end{align}
where $Noise\textunderscore Energy$ is the total energy that spent in generating the noise sequences during an attack which is the sum of the energy of all noise sequences that are injected into the leakage traces that are collected by the attacker. $Noise\textunderscore Energy = \sum_{i=1}^{I}\text{E}[{|N_{r}^{(i)}|^2}]$~(or $\sum_{i=1}^{I}\text{E}[|\widehat{N_{a}}^{(i)}|^2$), where $N_{r}^{(i)}$~($\widehat{N_{a}}^{(i)}$) is the random noise sequence~(artificial noise sequence) that has been added to the $i$-th leakage trace and $I$ is the total number of the leakage traces. ${Successful\textunderscore Key\textunderscore Recovery}$ is a binary value, which indicates whether the attacker is able to recover the secret value during an attack. If the attacker recovers the key, the value of EE is zero; otherwise, the larger the EE is, the less energy has been spent to cause an unsuccessful attack, in other words, it is more energy efficient. 

\subsection{Security Analysis}
As discussed in Sec.~\ref{sec:random_noise}, with the injection of the random noise, the required number of leakage traces~(quires) $I_q$ is highly increased, in order to achieve a target $P_s$ during the attack. More specifically, given the following inequality derived from~\cite{eloi2019best} 
\begin{align}
I_q \geq \frac{B+(P_s -1)\log(2^B -1)-H(P_s) }{\frac{1}{2}\log(1+\text{SNR})}~~, \nonumber
\end{align}
as an example, if the \text{SNR} is decreased from 0.04 to 0.01, in order to obtain same $P_s$, the minimum number of traces $I^{\text{min}}_{q}$ which equals to the lower bound is increased by near a factor of 4, which means much more numbers of attack queries are required. This gives the basis for enhancing the security protection of the system under SCAs. Besides, we set the function:~$\xi (P_e) = B+(P_s -1)\log(2^B -1)-H(P_s)$ for simplicity in the following.  

Furthermore, same as in~\cite{eloi2019best}, we also use the Kullback–Leibler divergence to analyze the relative entropy between any two distributions under additional noise injection, for getting a more tight bound on $P_s$:~$\xi (P_e)$. Suppose we have $\mathbb{P}_{({\boldsymbol{L}_j}|K_i,\textbf{T})}$ with~($i=1,2$) follows a multivariate normal distribution:~$\mathcal{N}( \boldsymbol{Y(X)}_j, \sigma^2_N \boldsymbol{I})$, where $\boldsymbol{Y(X)}_j=f(\phi(K_i, \textbf{T}))_j=(f(\phi(K_i, {T}_1))_j,\dots,f(\phi(K_i, {T}_{I_q}))_j)$, $f(\cdot)_j$ is the $j$-th point of the leakage function vector~($j=1\dots,m$), and $\boldsymbol{I}$ is an identity matrix. For any two keys $K_1$ and $K_2$, the Kullback-Leibler divergence on their distributions at $j$-th point is:
\begin{align}
\text{D}_j\left(\mathbb{P}_{({\boldsymbol{L}_j}|K_1,\textbf{T})}||\mathbb{P}_{({\boldsymbol{L}_j}|K_2,\textbf{T})}\right)=\frac{\| f(\phi(K_1, \textbf{T}))_j - f(\phi(K_2, \textbf{T}))_j \|^2_2}{2\sigma_N^2}~~, \label{eq181}
\end{align}
with $j=1,\dots,m$. We can easily find with the injection of noise sequence, $\sigma_N^2$ is increased to $\sigma^2_N + \sigma^2_a$. Hence the divergence is therefore decreasing, i.e., by setting $\sigma_a=\sigma_N$, the divergence value is reduced by a factor of 2. Based on the main result~(inequality (5)) in~\cite{eloi2019best}, here we have the following inequality holds:
\begin{align}
\xi (P_e) \leq \min_{j} -\mathbb{E}_{\textbf{T}}\mathbb{E}_{K_1}\log\mathbb{E}_{K_2}e^{-\text{D}_j}~~, \nonumber
\end{align}
where $\text{D}_j$ is defined in Eq.~\eqref{eq181}. It's easily to find $\xi (P_e) $ which is the LHS of the inequality, is strictly increasing with the increase of $P_s$. When $\textbf{T}$~(also $I_q$) is fixed, the upper bound~(the RHS) is decreasing with the injection of additional noise with $\sigma_a$. This decreased upper bound therefore downgrades the target $P_s$, which means, in another way, the effectiveness of defense is enhanced. 


We further illustrate our proposed mechanism can provide same security as the random noise generation but in energy efficient manner. Let's denote $\Omega_m$ as the set of the indices of the secret-dependent samples among the leakage trace under a given target implementation. $\Omega_{F_o}$ is the set of the indices obtained by our approach where $\Omega_{F_o} \subseteq \Omega_m$. It's obviously that the SNR of the recorded leakage sample point is meaningful only if this point is in set $\Omega_m$~(means the Eq.~\eqref{eq181} is valid under this case). Therefore the majority~($\geq 95\%$ in some cases) samples (they are in the complement of $\Omega_m$:~$\Omega_m^{\complement}$) covered by the sequence under random noise generation are nothing but independent noise, which don't contain any useful information. As a subset of $\Omega_m$, $\Omega_{F_o}$ only contains the useful leakage samples. When $\Omega_{F_o}=\Omega_m $, our mechanism obtains the same power as random noise generation with much less energy spent. Besides, if our mechanism gets same amount of energy budget as the random noise generation, our mechanism can provide much more stronger security than random noise generation by allocating the extra energy on the variance or gain of the impulsive noise samples in $\Omega_{F_o}$.       

Besides, our proposed scheme is compatible to work together with the masking~\cite{prouff2013masking}, for further enhancing the system's security. As derived in~\cite{ito2022success}, given the number of masking shares is $d$, $\xi (P_e)$ is upper-bounded by
\begin{align}
\xi (P_e) \leq I_q \log \left((2^B-1)(2\ln2)^d\prod^{d}_{i=1}MI(L^{s}_i;S_i)+1\right)~~, \label{eq182}
\end{align}
where $S_i$ is the $i$-th share which results in the leakage trace $L^{s}_i$. Since each leakage $L^{s}_i$ is only dependent on $S_i$ and is independent to other shares, we can treat each $MI(L^{s}_i;S_i)$ as an independent channel and model it as the noisy channel as Eq.~\eqref{eq1}. By applying our noise generation mechanism on the leakage trace from each share, the mutual information on each share's channel is further decreased:~Let's assume the MI value on each channel is decreased by a factor of $\triangle_a$ caused by the impulsive noise sequence with the variance of $\sigma_a$, then the value of the continued product among all MI is decreased by a factor of $\triangle^d_a$. Based on Eq.~\eqref{eq182}, $\xi (P_e) $ is therefore largely decreased~(so is the $P_e$), under fixed $I_q$. 

\subsection{Time Delay}
From Sec.~\ref{sec:energy-efficient}, it can be found that the effectiveness of the scheme depends on the alignment of the timing of impulsive noise and the target segment in the leakage trace. Ideally, the noise samples generated by Eq.~\eqref{eq17} are expected to be exactly overlapped with the leakage samples in the same or close approximate timing window. However, in practice, due to the likely delay introduced by hardware circuit, the generated noise impulse sequence may not be perfectly aligned with the leakage trace, for example, the starting point of two sequences is different. This is also called {\em asynchronization}. Our result is derived under synchronized assumption, which can be viewed as the theoretical boundary in an ideal situation. 

\section{Experimental Results}
\label{sec:experimental-results}
We evaluate our proposed scheme on the Grizzly benchmark dataset described in~\cite{computer2013}.
Grizzly is based on leakage data collected from the data bus of the Atmel XMEGA 256 A3U, which is a widely used 8-bit micro-controller. 
Given the 8-bit nature of the system, there are 256 keys. For each key $K \in \{0, 1, \ldots, 255\}$, 3072 raw traces are recorded. Note since the target micro-controller only executes the load instruction on each key, we just treat $\text{T}$ is fixed as constant during the operations. Hence now we have $X=K$. Typically, these recorded traces are divided into two sets: a \emph{profiling} set and an \emph{attack} set. Each raw trace has 2500 samples, and each sample represents the total current consumption in all CPU ground pins.

The compression methods that we used in the experiments include 3ppc, 20ppc, allap, LDA, and PCA, which are commonly used compression methods in side-channel attacks. We pick 3ppc, 20ppc, LDA, and PCA as the candidates of the compression methods used in~\textbf{Algorithm~1}. In the Grizzly dataset, the number of samples obtained by these sample selection methods in each trace typically are:~$18\sim30$ for 3ppc, $75\sim79$ for 20ppc, and $125$ for allap, respectively~\cite{choudary2013efficient}. For PCA and LDA, we set the dimension after compression as 4~(now $m_c=4$) during the system design phase. 

We compare our noise generation scheme to two other methods:~Random Full Sequence and Random Partial Sequence. Random Full Sequence~\cite{protect2000shamir} generates random noise at all time. As a result, all leakage samples are covered by random noise samples. 
Random Partial Sequence generates noises for adding onto only a randomly selected subset of the raw leakage samples. In order to fairly compare Random Partial Sequence to our scheme, we have both methods generate the same number $|\Omega_{F_o}|$ of noise samples. Hence the two scheme only differ in the selection of raw leakage samples to inject the noise into. In the following, we use the term $\textbf{PrArN}$ to denote \textbf{our} proposed practical artificial noise generation scheme, the term $\textbf{SArN}$ for the previous sample selection based noise generation scheme~\cite{Jin2022}, and the terms $\textbf{RnF}$ and $\textbf{RnP}$ for Random Full Sequence and Random Partial Sequence, respectively.  

\subsection{Metrics and Settings}
During the attack phase, for each key $k$, we independently run the attack 300 times by randomly picking the leakage traces from the \emph{attack} set. We use the ${Successful\textunderscore Recovery\textunderscore Rate}$~($SRR$) to measure the attack, which is the average success  probability over \textbf{all} keys:
\begin{align}
{Successful\textunderscore Recovery\textunderscore Rate}=\frac{\sum_{K=0}^{2^B-1}{P}_{attack}(K)}{2^B}~~, \nonumber
\end{align}
where ${{P}_{attack}(K)}=\frac{{N_{hit}(K)}}{{N_T}}$. Here ${N_T}$ is the number of tests~(in our case, $N_T$=300) and ${N_{hit}(k)}$ stands for how many times that key $k$ is successfully recovered during all ${N_T}$ tests. We defined the EE in Eq.~\eqref{eq18}. In our experiments, we use $EE_{avg}$ to measure the average energy efficiency of the noise generation for \textbf{all} keys, which is: 
\begin{align}
EE_{avg}=\frac{\sum_{K=0}^{2^B-1}\frac{\sum_{t=1}^{N_T}EE(K, t)}{N_T}}{2^B}~~, \nonumber
\end{align}
where $EE(K, t)$ is the $EE$ value for an attack on key $K$ in $t$-th test. To scale the result, we normalize the $Noise\textunderscore Energy$ in Eq.~\eqref{eq18} by letting $Noise\textunderscore Energy = \frac{Noise\textunderscore Energy}{N_T N_s \sigma^2}$, where $\sigma$ is the variance of the generated noise sequence, and $N_s$ is the number of sample per trace. In order to fairly compare all noise generation methods, both $N_r$ and $N_a$ are generated from the same distribution. The parameters of the noise distribution is obtained from the raw leakage trace:~We collect the raw leakage traces under $k = 0$ and compute the mean and variance. 
We also use 1000 traces to compute the matrix $P$ for any given compression method. 

\subsection{Experimental Data and Analysis}
In the following, we will use $I_a$ and $I_p$ to denote the number of profiling and attack traces for each key, respectively. 

\subsubsection{Results on Standard Template Attack}
The attack algorithm used in the experiments is the linear model based Template Attack described in~\cite{schindler2005stochastic,Jin2018}. In this attack, the leakage model is represented by a linear model (also called {\em stochastic model} in this context), and during profiling the leakage model is estimated from all or a subset of keys. Based on the estimated leakage model, the template, which typically is in form of Gaussian Model, for each key is built. In the attack phase, we select the candidate key with the maximum likelihood in matching. In our experiments, we use all keys~(256 keys) for profiling the leakage model. As a baseline for key recovery, we also present the results of the original Template Attack (denoted by \textbf{OA}), which is the version without noise generated. Note as there is no noise generated for \textbf{OA}, EE has no meaning for \textbf{OA}.

\begin{figure}

      \centering
      \begin{subfigure}[t]{0.49\linewidth}
          \centering
          \includegraphics[width = 1.02\linewidth]{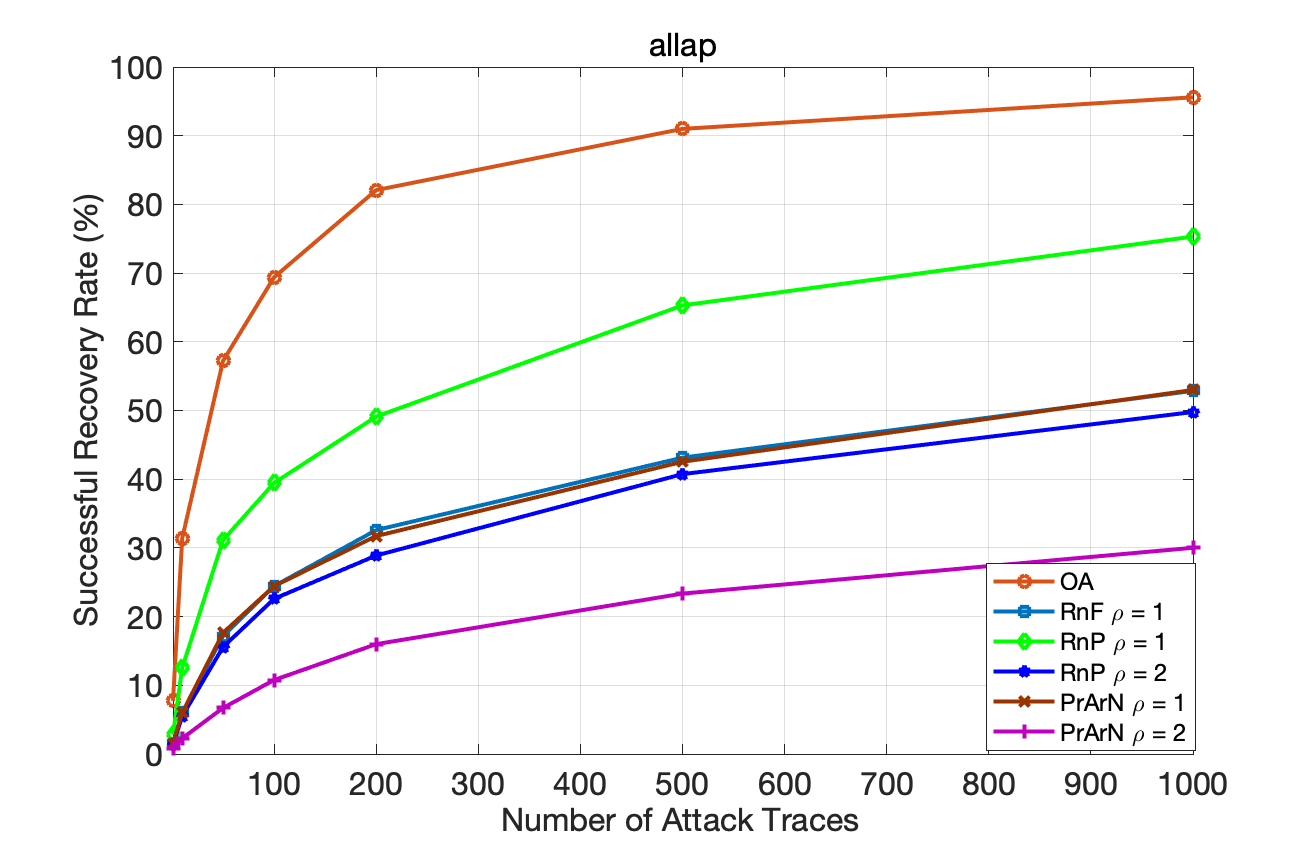}
          \caption{{\em SRR}, allap}
          \label{fig3_1}
      \end{subfigure}
      \hfill
      \begin{subfigure}[t]{0.49\linewidth}
          \centering
          \includegraphics[width = 1.02\linewidth]{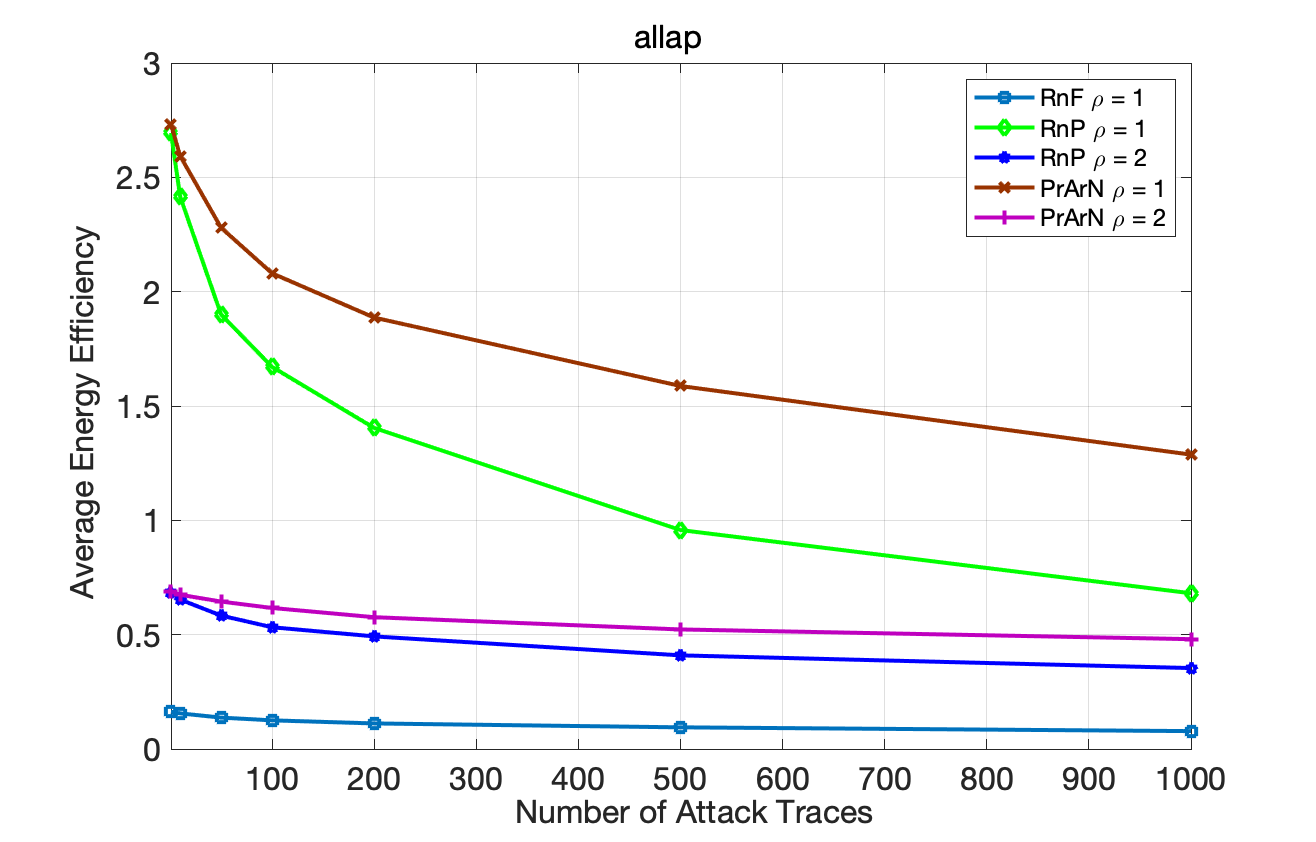}
          \caption{{$EE_{avg}$}, allap}
          \label{fig3_2}
      \end{subfigure}
      \hfill
    \begin{subfigure}[t]{0.49\linewidth}
          \centering
          \includegraphics[width = 1.02\linewidth]{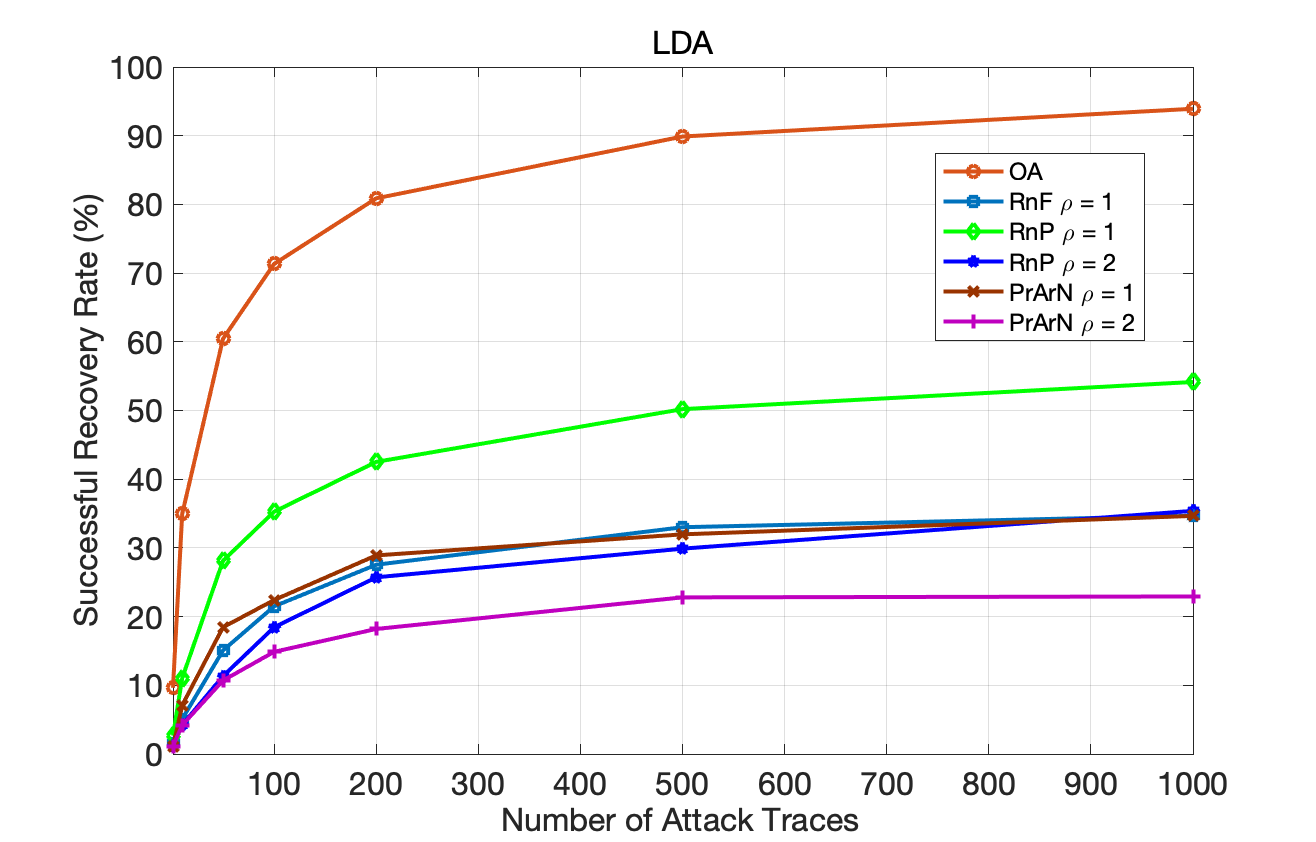}
          \caption{{\em SRR}, LDA}
          \label{fig3_3}
      \end{subfigure}
      \hfill
      \begin{subfigure}[t]{0.49\linewidth}
          \centering
          \includegraphics[width = 1.02\linewidth]{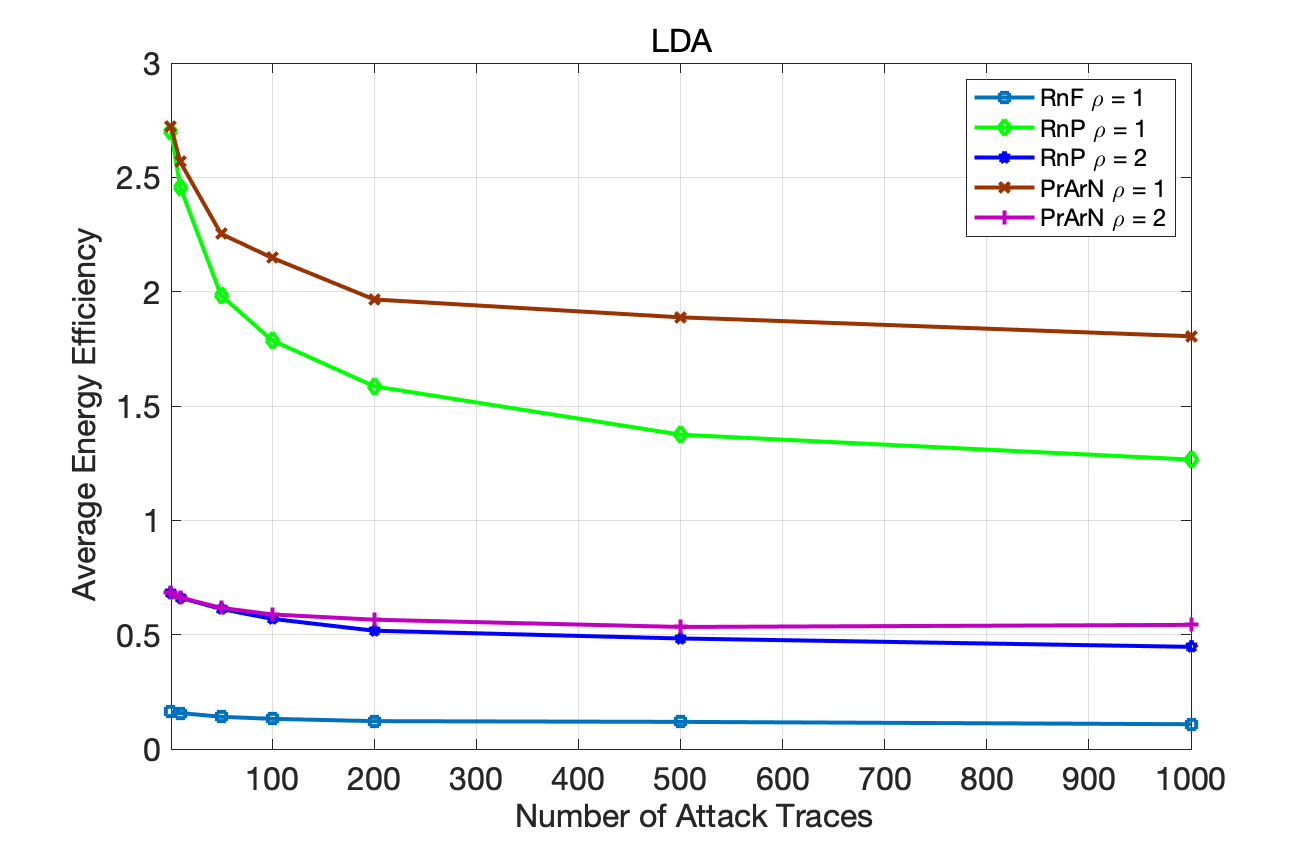}
          \caption{{$EE_{avg}$}, LDA}
          \label{fig3_4}
      \end{subfigure}
     \caption{Probability of Successful Recovery, and Average EE, versus Different Attack Traces $I_a$, under Compression Methods:~allap and LDA for Attacker, respectively}
         \label{fige1}
\end{figure}

Fig.~\ref{fige1} shows the recovery rates and EE values with different numbers of $I_a$ under attack compression methods:~allp and LDA, respectively. Here $I_p$=500, and $I_a$ varies from 1 to 1000. $R_A$ is set to 150. We also compare the effect of the gain by presenting the results with the increased gain $\rho=2$ for $\textbf{PrArN}$ and $\textbf{RnP}$~(now $E_A^{'}=4E_A$). From the figure, in general, it shows the proposed method \textbf{PrArN} outperforms all other noise generation methods for combating the side-channel attacks in both effectiveness~(resistance) and efficiency~(EE). When evaluating the effectiveness~(Fig.~\ref{fig3_1}~and Fig.~\ref{fig3_3}), we can find when $\rho =1$, even when $I_a$=1000, $\textbf{ArN}$ can still decrease the recovery rate from $\geq 93\%$~(\textbf{OA}'s number) to $53.41\%$~(allap's number) and $34.66\%$~(LDA's number), which is significant. Besides, the effectiveness of \textbf{PrArN} is always better than \textbf{RnP} with same gain factor. Considering both $\textbf{RnP}$ and $\textbf{PrArN}$ generate the same number of noise samples, this is a good proof that the effectiveness of our proposed model relies on the fact that the generation of the noise samples is artificially designed and particularly target on the useful leakage samples. Hence, it's also straightforward that $\textbf{RnP}$ 's EE values are always worse than $\textbf{PrArN}$'s, which is reflected in the experimental results in~(Fig.~\ref{fig3_2}~and Fig.~\ref{fig3_4}).

We also notice when an appropriate number of attack traces is used~(such as $I_a \geq 500$), both $\textbf{PrArN}$~($\rho=1$) and $\textbf{RnF}$ allow for almost the same recovery rates. If now we evaluate the efficiency by observing the EE values, we can clearly find that $\textbf{PrArN}$ always displays better EE value. As a result, it proves that to achieve \emph{same} or \emph{approximate} level of security, our method has the benefits in spending much less energy compared to the $\textbf{RnF}$. 
When increasing $\rho$=2, $\textbf{PrArN}$ further strongly enhances its power in defending SCAs. For example, when $I_a$=1000, \textbf{PrArN} can decrease the recovery rate to $30.52\%$~(allap) and $22.93\%$~(LDA), with average EE of 0.4803~(allap) and 0.5428~(LDA). As a baseline, the numbers of \textbf{RnF} with $\rho$=1 are:~$52.86\%$~(allap) and $34.68\%$~(LDA) in $SRR$, and 0.0782~(allap) and 0.1083~(LDA) in $EE_{avg}$. Compared to \textbf{RnF}, the energy efficiency improvement conducted by \textbf{PrArN} is tremendous. 

\begin{figure}
      \centering
      \begin{subfigure}[t]{0.49\linewidth}
          \centering
          \includegraphics[width = 1.02\linewidth]{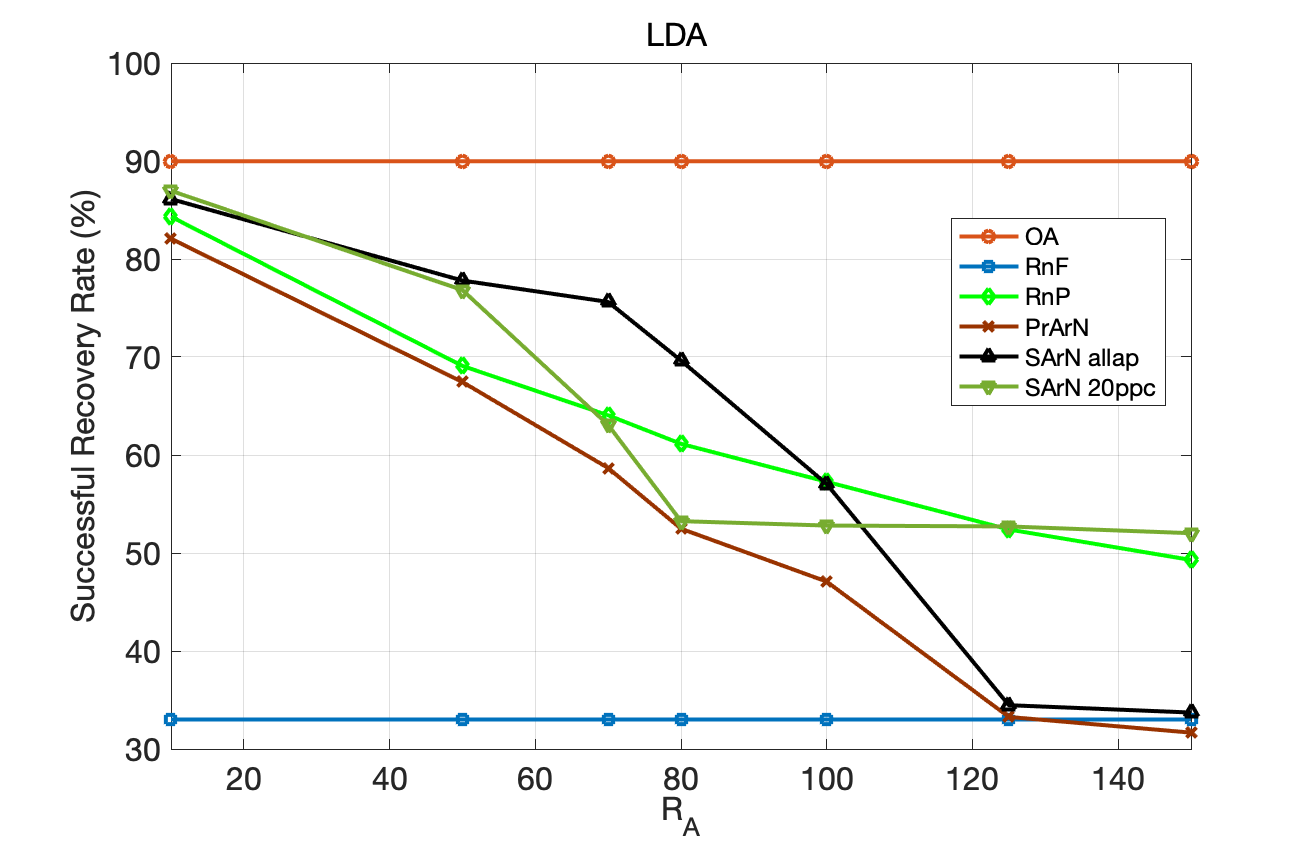}
          \caption{{\em SRR}, LDA}
          \label{fig2_1}
      \end{subfigure}
      \hfill
      \begin{subfigure}[t]{0.49\linewidth}
          \centering
          \includegraphics[width = 1.02\linewidth]{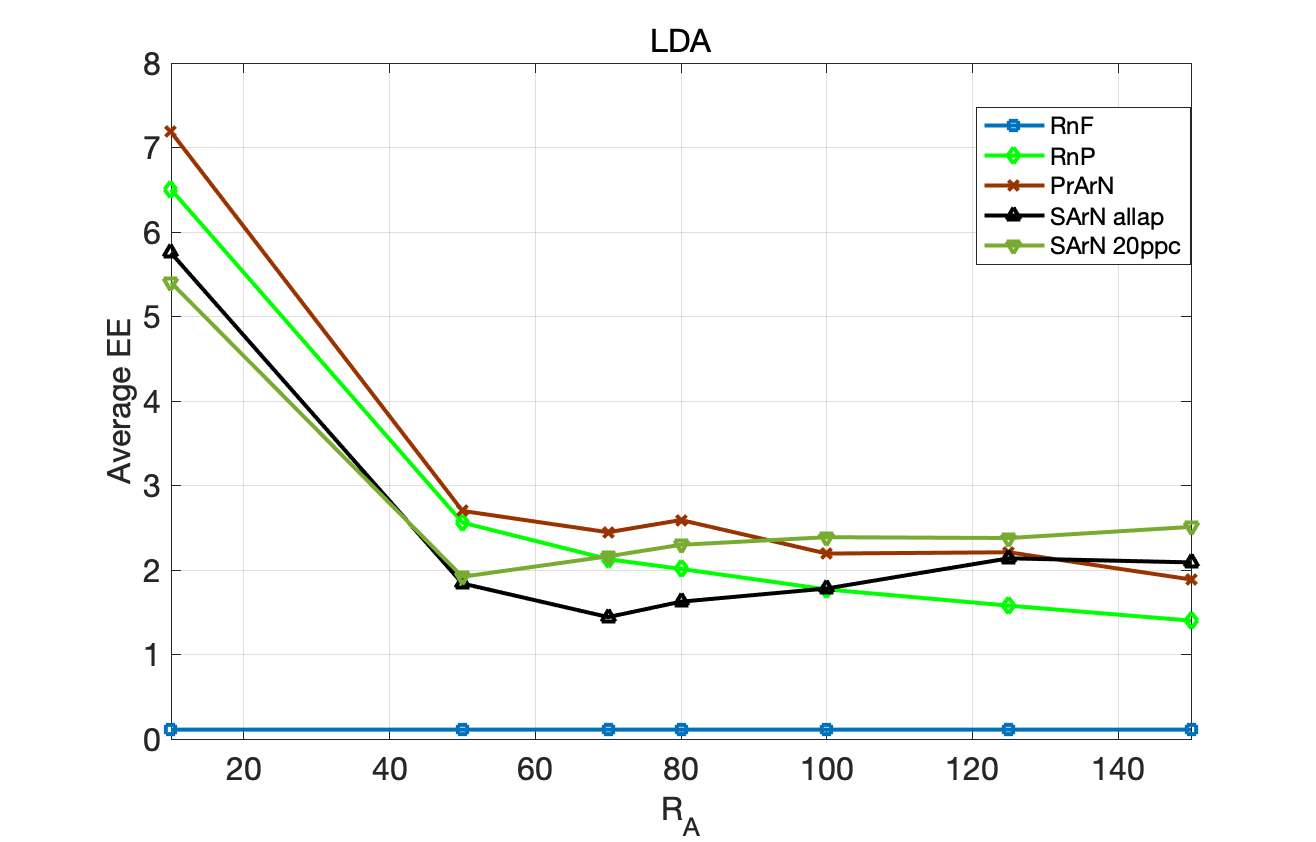}
          \caption{{$EE_{avg}$}, LDA}
          \label{fig2_2}
      \end{subfigure}

     \caption{Probability of Successful Recovery, and Average EE, versus Different $R_A$, under Compression Method:~LDA for Attacker}
         \label{fige2}

 \end{figure}

In Fig.~\ref{fige2}, we present the results of the successful recovery rate and the average EE for different values of $R_A$, under LDA as the compression method chosen by the attacker. Here $I_a$ and $I_p$ are all fixed to 500. We also fix the gain $\rho$ =1 to control the overhead of the noise sequences. Note as~\textbf{OA} has no noise generated, and $\textbf{RnF}$ always covers the full size of leakage trace, $R_A$ has no impact on them. We just repeat the experiments and take average on all $R_A$. 

In order to compare our model with the sample selection based method in~\cite{Jin2022}, we also present the results of \textbf{SArN} under both 20ppc and allap, which 
are good examples to prove the effectiveness of our proposed method. As we mentioned before, the solution in~\cite{Jin2022} is in fact a special case of our solution. For \textbf{SArN}~(20ppc), in Grizzly dataset, only when $A \geq 80$, it can get the full solution of $F$~(for allap this number is 125). When $A < 80$, \textbf{PrArN} outperforms \textbf{SArN}~(both 20ppc and allap) in both effectiveness and efficiency. This is because our designed approach, by aggregating the noise generation patterns from a variety of different compression methods, is much more effective in identifying the most significant useful leakage samples, by selecting them through the ordering of their importance~(Eq.~\eqref{eq17}). On the contrary, the sample selection based methods only relies on a single category of compression methods, which highly limits the chance of extracting the useful samples by exploiting the information in other domains. Furthermore, the sample selection based methods just randomly select a subset from the candidates~(the set of indices of "1"s), which would cause the problem that those most important samples can not be guaranteed to be selected by every time, especially when $R_A$ is small. This is also proved by the experimental results that when $R_A$ didn't achieve the full solution, the performance of the sample selection based methods is much worse compared to \textbf{PrArN}. For example when $R_A = 70$, the $SRR$ and $EE_{avg}$ for \textbf{PrArN} are $58.67\%$ and 2.4467, while for \textbf{SArN}~(allap) those are $75.62\%$ and 1.4427. When the full solution is guaranteed, \textbf{SArN} can obtain the optimal result in effectiveness, same as \textbf{PrArN}. Note the comparison of the EE between \textbf{PrArN} and \textbf{SArN} is only meaningful when both of them are under the same valid number of $R_A$, as \textbf{SArN} will only generate $|\Omega_{\hat{P}}|$ impulsive noises when $R_A$ is larger than $|\Omega_{\hat{P}}|$. 

\subsubsection{Results on A DPA Style based Attack}
Same as the work in~\cite{Choudary2018portablel}, we also use the same Grizzly
data set to simulate a DPA~(Differential Power Analysis) style~\cite{whitnall2015robust} based template attack on AES. Specifically, we treat each key $K$ as the actual output of $v=S(\text{T} \oplus K)$:~We first generate a random key byte $K^*$ as the target key. Then for each leakage trace, we can compute the corresponding plaintext byte \text{T} to make $v=S(\text{T} \oplus K^*)$ is the byte that actually be loaded and results in the leakage trace. During the attack phase, to each leakage trace $L_i$, the corresponding plaintext byte \text{T} is known. Then under this plaintext, for each candidate key byte $K$, we could compute the output of the Sbox operation:~$S(\text{T} \oplus K)$ and obtain the corresponding output value $v$. Therefore, by combining multiple attack traces, the linear discriminant of each candidate key byte $K$ can be computed, as presented in~\cite{Choudary2018portablel}. Eventually, we guess the target key by picking the key byte which brings the maximum linear discriminant value. Note since now our target is just a \emph{single} key byte, both the $SRR$ and $EE_{avg}$ are the results that over one key. 
Besides, in order to more fairly compare \textbf{PrArN} with \textbf{SArN}, here we make a modification on the original \textbf{SArN}:~when $R_A > |\Omega_{\hat{P}}|$, now \textbf{SArN} will also generate $R_A$ impulsive noises, but these additional $R_A - |\Omega_{\hat{P}}|$ noises are randomly generated and added on the lekage trace.  

Tab.~\ref{tab1} shows the successful recovery rate and the average EE for different methods~$(P_A$) chosen by attacker: LDA and allap. We fixed $I_p = 500$ and $I_a = 100$, $R_A=125$, and $\rho = 1$. \textbf{SArN} is under 20ppc. From the table, we can still easily find that our methods outperforms \textbf{RnP} and \textbf{SArN} in both $SRR$ and $EE_{avg}$, and \textbf{RnF} in $EE_{avg}$, under all two cases. For example, when $P_A$=allap, \textbf{PrArN} brings near $10\%$ decreasing on the recovery rate compared with \textbf{SArN}. Compared to \textbf{RnF}, \textbf{PrArN} obtains almost same recovery rate but with near 20 times gain in energy efficiency. Moreover, even now $R_A=125$, the original \textbf{SArN}~(20ppc) can only generate a smaller number of impulsive noises. Even now we make up the differences on the number of noises, as \textbf{SArN} has no knowledge about other useful sample points that are not belonging to the set $|\Omega_{\hat{P}}|$, hence only random noises are generated, which is far less effective to resistant the SCAs. This also shows the limitation of \textbf{SArN}, and also a good proof of superior effectiveness of our proposal which makes the best effort to target on these useful information under a given energy budget in noise generation. 

\begin{table*}[!htbp]
\centering

{\small
\begin{tabular}{|c|c|c|c|c|c|c|c|c|c|}
\hline
\multirow{2}*{\diagbox{$P_A$}{Methods}}& OA & \multicolumn{2}{c|}{RnF} & \multicolumn{2}{c|}{RnP}  & \multicolumn{2}{c|}{SArN}  & \multicolumn{2}{c|}{PrArN} \\
\cline{2-10}
{}&{$SRR$~(\%)}&{$SRR$~(\%)}&{$EE_{avg}$}&{$SRR$~(\%)}&{$EE_{avg}$}&{$SRR$~(\%)}&{$EE_{avg}$}&{$SRR$~(\%)}&{$EE_{avg}$}\\
\hline
allap&{100.00}&{\textbf{58.83}}&{0.0686}&{80.87}&{0.6792}&{68.13}&{1.0521} &{\textbf{58.47}}&{\textbf{1.3885}}\\
\hline
LDA&{100.00}&{\textbf{45.27}}&{0.1028}&{62.39}&{1.2489}&{56.72}&{1.4799} &{\textbf{44.93}}&{\textbf{1.8842}}\\
\hline
\end{tabular}
}
\caption{Probability of Successful Recovery, and Average EE, versus Different Compression Methods for Attacker}\label{tab1}
\end{table*}

\section{Conclusion}
\label{sec:conclusion}
In this paper, we propose a practical approach for generating energy efficient artificial noise sequences to prevent SCAs. Our design of the artificial noise is the impulsive noise sequence, which exploits the sparsity among the raw leakage traces that only very few number of leakage samples contain the useful leakage information. Typically, the useful leakage information is extracted by the compression methods. Compared to previous work which only consider the sample selection based compression, our solution works for all categories of compression methods. Then by using channel capacity to measure the mutual information between the side-channel leakage and the secret, the optimal noise sequence design problem is translated to a side-channel's channel capacity minimization problem, and we obtain the optimum under a given energy budget. 

Furthermore, we also propose an approach to design a practical energy efficient noise generation mechanism by aggregating the optimal noise generation patterns under different compression methods. Compared to the previous noise generation which is only based on a specific compression method at a time, our approach enhances the power of protection and also the practicability. 


\bibliographystyle{ACM-Reference-Format}
\bibliography{MyRef.bib}


\begin{thebibliography}{40}


\ifx \showCODEN    \undefined \def \showCODEN     #1{\unskip}     \fi
\ifx \showDOI      \undefined \def \showDOI       #1{#1}\fi
\ifx \showISBNx    \undefined \def \showISBNx     #1{\unskip}     \fi
\ifx \showISBNxiii \undefined \def \showISBNxiii  #1{\unskip}     \fi
\ifx \showISSN     \undefined \def \showISSN      #1{\unskip}     \fi
\ifx \showLCCN     \undefined \def \showLCCN      #1{\unskip}     \fi
\ifx \shownote     \undefined \def \shownote      #1{#1}          \fi
\ifx \showarticletitle \undefined \def \showarticletitle #1{#1}   \fi
\ifx \showURL      \undefined \def \showURL       {\relax}        \fi
\providecommand\bibfield[2]{#2}
\providecommand\bibinfo[2]{#2}
\providecommand\natexlab[1]{#1}
\providecommand\showeprint[2][]{arXiv:#2}

\bibitem[Archambeau et~al\mbox{.}(2006)]%
        {Archambeau2006subspace}
\bibfield{author}{\bibinfo{person}{C{\'e}dric Archambeau},
  \bibinfo{person}{Eric Peeters}, \bibinfo{person}{Fran{\c{c}}ois-Xavier
  Standaert}, {and} \bibinfo{person}{Jean-Jacques Quisquater}.}
  \bibinfo{year}{2006}\natexlab{}.
\newblock \showarticletitle{{Template Attacks in Principal Subspaces}}. In
  \bibinfo{booktitle}{\emph{International Workshop on Cryptographic Hardware
  and Embedded Systems~(CHES'06)}}. Springer, \bibinfo{pages}{1--14}.
\newblock


\bibitem[Benadjila et~al\mbox{.}(2020)]%
        {ryad2020deep}
\bibfield{author}{\bibinfo{person}{Ryad Benadjila}, \bibinfo{person}{Emmanuel
  Prouff}, \bibinfo{person}{Remi Strullu}, \bibinfo{person}{Eleonora Cagli},
  {and} \bibinfo{person}{Cecile Dumas}.} \bibinfo{year}{2020}\natexlab{}.
\newblock \showarticletitle{{Deep Learning for Side-Channel Analysis and
  Introduction to ASCAD Database}}.
\newblock \bibinfo{journal}{\emph{Journal of Cryptographic Engineering}}
  \bibinfo{volume}{10} (\bibinfo{year}{2020}), \bibinfo{pages}{163--188}.
\newblock


\bibitem[Brier et~al\mbox{.}(2004)]%
        {brier2004correlation}
\bibfield{author}{\bibinfo{person}{Eric Brier}, \bibinfo{person}{Christophe
  Clavier}, {and} \bibinfo{person}{Francis Olivier}.}
  \bibinfo{year}{2004}\natexlab{}.
\newblock \showarticletitle{{Correlation Power Analysis with a Leakage Model}}.
  In \bibinfo{booktitle}{\emph{International Workshop on Cryptographic Hardware
  and Embedded Systems~(CHES'04)}}. Springer, \bibinfo{pages}{16--29}.
\newblock


\bibitem[Chari et~al\mbox{.}(2003)]%
        {chari2003template}
\bibfield{author}{\bibinfo{person}{S. Chari}, \bibinfo{person}{J.~R Rao}, {and}
  \bibinfo{person}{P. Rohatgi}.} \bibinfo{year}{2003}\natexlab{}.
\newblock \showarticletitle{{Template Attacks}}. In
  \bibinfo{booktitle}{\emph{International Workshop on Cryptographic Hardware
  and Embedded Systems~(CHES'03)}}. \bibinfo{publisher}{Springer},
  \bibinfo{pages}{13--28}.
\newblock


\bibitem[Cheng et~al\mbox{.}(2022)]%
        {cheng22masked}
\bibfield{author}{\bibinfo{person}{Wei Cheng}, \bibinfo{person}{Yi Liu},
  \bibinfo{person}{Sylvain Guilley}, {and} \bibinfo{person}{Olivier Rioul}.}
  \bibinfo{year}{2022}\natexlab{}.
\newblock \showarticletitle{{Attacking Masked Cryptographic Implementations:
  Information-Theoretic Bounds}}. In \bibinfo{booktitle}{\emph{2022 IEEE
  International Symposium on Information Theory (ISIT'22)}}.
  \bibinfo{publisher}{IEEE}, \bibinfo{pages}{654--659}.
\newblock


\bibitem[Choudary and Kuhn(2013)]%
        {choudary2013efficient}
\bibfield{author}{\bibinfo{person}{Marios~O. Choudary} {and}
  \bibinfo{person}{Markus~G. Kuhn}.} \bibinfo{year}{2013}\natexlab{}.
\newblock \showarticletitle{{Efficient template attacks}}. In
  \bibinfo{booktitle}{\emph{International Conference on Smart Card Research and
  Advanced Applications~(CARDIS'13)}}. \bibinfo{publisher}{Springer},
  \bibinfo{pages}{253--270}.
\newblock


\bibitem[Choudary and Kuhn(2018)]%
        {Choudary2018portablel}
\bibfield{author}{\bibinfo{person}{Marios~O. Choudary} {and}
  \bibinfo{person}{Markus~G. Kuhn}.} \bibinfo{year}{2018}\natexlab{}.
\newblock \showarticletitle{{Efficient, Portable Template Attacks}}.
\newblock \bibinfo{journal}{\emph{IEEE Transactions on Information Forensics
  and Security}} \bibinfo{volume}{13}, \bibinfo{number}{2}
  (\bibinfo{year}{2018}), \bibinfo{pages}{490--501}.
\newblock


\bibitem[Chérisey et~al\mbox{.}(2019)]%
        {Cherisey2019information}
\bibfield{author}{\bibinfo{person}{Éloi~de Chérisey},
  \bibinfo{person}{Sylvain Guilley}, \bibinfo{person}{Olivier Rioul}, {and}
  \bibinfo{person}{Pablo Piantanida}.} \bibinfo{year}{2019}\natexlab{}.
\newblock \showarticletitle{{An Information-Theoretic Model for Side-Channel
  Attacks in Embedded Hardware}}. In \bibinfo{booktitle}{\emph{2019 IEEE
  International Symposium on Information Theory~(ISIT'19)}}.
  \bibinfo{publisher}{IEEE}, \bibinfo{pages}{310--315}.
\newblock


\bibitem[Cover and Thomas(2006)]%
        {elements2006cover}
\bibfield{author}{\bibinfo{person}{Thomas~M. Cover} {and}
  \bibinfo{person}{Joy~A. Thomas}.} \bibinfo{year}{2006}\natexlab{}.
\newblock \bibinfo{booktitle}{\emph{{Elements of Information Theory}}}.
\newblock \bibinfo{publisher}{John Wiley \& Sons, Inc.}
\newblock
\showISBNx{0471241959}


\bibitem[Das et~al\mbox{.}(2019)]%
        {das2019cross}
\bibfield{author}{\bibinfo{person}{Debayan Das}, \bibinfo{person}{Anupam
  Golder}, \bibinfo{person}{Josef Danial}, \bibinfo{person}{Santosh Ghosh},
  \bibinfo{person}{Arijit Raychowdhury}, {and} \bibinfo{person}{Shreyas Sen}.}
  \bibinfo{year}{2019}\natexlab{}.
\newblock \showarticletitle{{X-DeepSCA: Cross-Device Deep Learning Side Channel
  Attack}}. In \bibinfo{booktitle}{\emph{2019 56th ACM/IEEE Design Automation
  Conference (DAC'19)}}. \bibinfo{publisher}{ACM}, \bibinfo{pages}{1--6}.
\newblock


\bibitem[Das et~al\mbox{.}(2017)]%
        {das2017attenuated}
\bibfield{author}{\bibinfo{person}{Debayan Das}, \bibinfo{person}{Shovan
  Maity}, \bibinfo{person}{Saad~Bin Nasir}, \bibinfo{person}{Santosh Ghosh},
  \bibinfo{person}{Arijit Raychowdhury}, {and} \bibinfo{person}{Shreyas Sen}.}
  \bibinfo{year}{2017}\natexlab{}.
\newblock \showarticletitle{{High Efficiency Power Side-Channel Attack Immunity
  using Noise Injection in Attenuated Signature Domain}}. In
  \bibinfo{booktitle}{\emph{2017 IEEE International Symposium on Hardware
  Oriented Security and Trust~(HOST'17)}}. \bibinfo{publisher}{IEEE},
  \bibinfo{pages}{62--67}.
\newblock


\bibitem[de~Chérisey et~al\mbox{.}(2019)]%
        {eloi2019best}
\bibfield{author}{\bibinfo{person}{Eloi de Chérisey}, \bibinfo{person}{Sylvain
  Guilley}, \bibinfo{person}{Olivier Rioul}, {and} \bibinfo{person}{Pablo
  Piantanida}.} \bibinfo{year}{2019}\natexlab{}.
\newblock \showarticletitle{Best Information is Most Successful: Mutual
  Information and Success Rate in Side-Channel Analysis}. In
  \bibinfo{booktitle}{\emph{IACR Transactions on Cryptographic Hardware and
  Embedded Systems~(CHES'19)}}, Vol.~\bibinfo{volume}{2019}.
  \bibinfo{publisher}{Springer}, \bibinfo{pages}{49–79}.
\newblock


\bibitem[De~Cnudde et~al\mbox{.}(2016)]%
        {thomas2016masking}
\bibfield{author}{\bibinfo{person}{Thomas De~Cnudde}, \bibinfo{person}{Oscar
  Reparaz}, \bibinfo{person}{Beg{\"u}l Bilgin}, \bibinfo{person}{Svetla
  Nikova}, \bibinfo{person}{Ventzislav Nikov}, {and} \bibinfo{person}{Vincent
  Rijmen}.} \bibinfo{year}{2016}\natexlab{}.
\newblock \showarticletitle{{Masking AES with d+1 Shares in Hardware}}. In
  \bibinfo{booktitle}{\emph{International Conference on Cryptographic Hardware
  and Embedded Systems~(CHES'16)}}. \bibinfo{publisher}{Springer},
  \bibinfo{pages}{194--212}.
\newblock


\bibitem[Doget et~al\mbox{.}(2011)]%
        {doget2011univariate}
\bibfield{author}{\bibinfo{person}{Julien Doget}, \bibinfo{person}{Emmanuel
  Prouff}, \bibinfo{person}{Matthieu Rivain}, {and}
  \bibinfo{person}{Fran{\c{c}}ois-Xavier Standaert}.}
  \bibinfo{year}{2011}\natexlab{}.
\newblock \showarticletitle{{Univariate Side Channel Attacks and Leakage
  Modeling}}.
\newblock \bibinfo{journal}{\emph{Journal of Cryptographic Engineering}}
  \bibinfo{volume}{1}, \bibinfo{number}{2} (\bibinfo{year}{2011}),
  \bibinfo{pages}{123--144}.
\newblock


\bibitem[Fang et~al\mbox{.}(2022)]%
        {fang2022scas}
\bibfield{author}{\bibinfo{person}{Qiang Fang}, \bibinfo{person}{Longyang Lin},
  \bibinfo{person}{Yao~Zu Wong}, \bibinfo{person}{Hui Zhang}, {and}
  \bibinfo{person}{Massimo Alioto}.} \bibinfo{year}{2022}\natexlab{}.
\newblock \showarticletitle{{Side-Channel Attack Counteraction via Machine
  Learning-Targeted Power Compensation for Post-Silicon HW Security Patching}}.
  In \bibinfo{booktitle}{\emph{2022 IEEE International Solid-State Circuits
  Conference~(ISSCC'22)}}, Vol.~\bibinfo{volume}{65}.
  \bibinfo{publisher}{IEEE}, \bibinfo{pages}{1--3}.
\newblock


\bibitem[Faruque et~al\mbox{.}(2016)]%
        {faruque2016acoustic}
\bibfield{author}{\bibinfo{person}{Al Faruque}, \bibinfo{person}{Mohammad
  Abdullah}, \bibinfo{person}{Sujit~Rokka Chhetri}, \bibinfo{person}{Arquimedes
  Canedo}, {and} \bibinfo{person}{Jiang Wan}.} \bibinfo{year}{2016}\natexlab{}.
\newblock \showarticletitle{{Acoustic Side-Channel Attacks on Additive
  Manufacturing Systems}}. In \bibinfo{booktitle}{\emph{Proceedings of the 7th
  International Conference on Cyber-Physical Systems(ICCPS'16)}}. IEEE,
  \bibinfo{pages}{1--10}.
\newblock


\bibitem[Gandolfi et~al\mbox{.}(2001)]%
        {gandolfi2001electromagnetic}
\bibfield{author}{\bibinfo{person}{Karine Gandolfi},
  \bibinfo{person}{Christophe Mourtel}, {and} \bibinfo{person}{Francis
  Olivier}.} \bibinfo{year}{2001}\natexlab{}.
\newblock \showarticletitle{{Electromagnetic Analysis: Concrete Results}}. In
  \bibinfo{booktitle}{\emph{International Workshop on Cryptographic Hardware
  and Embedded Systems~(CHES'01)}}. Springer, \bibinfo{pages}{251--261}.
\newblock


\bibitem[Gierlichs et~al\mbox{.}(2006)]%
        {gierlichs2006templates}
\bibfield{author}{\bibinfo{person}{Benedikt Gierlichs},
  \bibinfo{person}{Kerstin Lemke-Rust}, {and} \bibinfo{person}{Christof Paar}.}
  \bibinfo{year}{2006}\natexlab{}.
\newblock \showarticletitle{{Templates vs. Stochastic Methods}}. In
  \bibinfo{booktitle}{\emph{International Workshop on Cryptographic Hardware
  and Embedded Systems~(CHES'06)}}. Springer, \bibinfo{pages}{15--29}.
\newblock


\bibitem[Goel and Negi(2008)]%
        {goel2008guarantee}
\bibfield{author}{\bibinfo{person}{Satashu Goel} {and} \bibinfo{person}{Rohit
  Negi}.} \bibinfo{year}{2008}\natexlab{}.
\newblock \showarticletitle{{Guaranteeing Secrecy using Artificial Noise}}.
\newblock \bibinfo{journal}{\emph{IEEE Transactions on Wireless
  Communications}} \bibinfo{volume}{7}, \bibinfo{number}{6}
  (\bibinfo{year}{2008}), \bibinfo{pages}{2180--2189}.
\newblock


\bibitem[Goldsmith et~al\mbox{.}(2003)]%
        {gold2003capacity}
\bibfield{author}{\bibinfo{person}{Andrea Goldsmith}, \bibinfo{person}{Syed~Ali
  Jafar}, \bibinfo{person}{Nihar Jindal}, {and} \bibinfo{person}{Sriram
  Vishwanath}.} \bibinfo{year}{2003}\natexlab{}.
\newblock \showarticletitle{{Capacity Limits of MIMO Channels}}.
\newblock \bibinfo{journal}{\emph{IEEE Journal on Selected Areas in
  Communications}} \bibinfo{volume}{21}, \bibinfo{number}{5}
  (\bibinfo{year}{2003}), \bibinfo{pages}{684--702}.
\newblock


\bibitem[Grizzly({[n.\,d.]})]%
        {computer2013}
\bibfield{author}{\bibinfo{person}{Grizzly}.}
  \bibinfo{year}{[n.\,d.]}\natexlab{}.
\newblock
  \showarticletitle{http://www.cl.cam.ac.uk/research/security/datasets/grizzly/}.
\newblock  (\bibinfo{year}{[n.\,d.]}).
\newblock


\bibitem[G{\"u}neysu and Moradi(2011)]%
        {generic2011tim}
\bibfield{author}{\bibinfo{person}{Tim G{\"u}neysu} {and} \bibinfo{person}{Amir
  Moradi}.} \bibinfo{year}{2011}\natexlab{}.
\newblock \showarticletitle{{Generic Side-Channel Countermeasures for
  Reconfigurable Devices}}. In \bibinfo{booktitle}{\emph{International Workshop
  on Cryptographic Hardware and Embedded Systems~(CHES'11)}}.
  \bibinfo{publisher}{Springer}, \bibinfo{pages}{33--48}.
\newblock


\bibitem[Heuser et~al\mbox{.}(2014)]%
        {heuser2014good}
\bibfield{author}{\bibinfo{person}{Annelie Heuser}, \bibinfo{person}{Olivier
  Rioul}, {and} \bibinfo{person}{Sylvain Guilley}.}
  \bibinfo{year}{2014}\natexlab{}.
\newblock \showarticletitle{{Good is Not Good Enough: Deriving Optimal
  Distinguishers from Communication Theory}}. In
  \bibinfo{booktitle}{\emph{International Workshop on Cryptographic Hardware
  and Embedded Systems~(CHES'14)}}. \bibinfo{publisher}{Springer},
  \bibinfo{pages}{55--74}.
\newblock


\bibitem[Ito et~al\mbox{.}(2022)]%
        {ito2022success}
\bibfield{author}{\bibinfo{person}{Akira Ito}, \bibinfo{person}{Rei Ueno},
  {and} \bibinfo{person}{Naofumi Homma}.} \bibinfo{year}{2022}\natexlab{}.
\newblock \showarticletitle{{On the Success Rate of Side-Channel Attacks on
  Masked Implementations: Information-Theoretical Bounds and Their Practical
  Usage}}. In \bibinfo{booktitle}{\emph{Proceedings of the 2022 ACM SIGSAC
  Conference on Computer and Communications Security~(CCS'22)}}.
  \bibinfo{publisher}{ACM}, \bibinfo{pages}{1521–1535}.
\newblock


\bibitem[Jin and Bettati(2018)]%
        {Jin2018}
\bibfield{author}{\bibinfo{person}{Shan Jin} {and} \bibinfo{person}{Riccardo
  Bettati}.} \bibinfo{year}{2018}\natexlab{}.
\newblock \showarticletitle{{Adaptive Channel Estimation in Side Channel
  Attacks}}. In \bibinfo{booktitle}{\emph{2018 IEEE International Workshop on
  Information Forensics and Security~(WIFS'18)}}. IEEE, \bibinfo{pages}{1--7}.
\newblock


\bibitem[Jin et~al\mbox{.}(2022)]%
        {Jin2022}
\bibfield{author}{\bibinfo{person}{Shan Jin}, \bibinfo{person}{Minghua Xu},
  \bibinfo{person}{Riccardo Bettati}, {and} \bibinfo{person}{Mihai
  Christodorescu}.} \bibinfo{year}{2022}\natexlab{}.
\newblock \showarticletitle{Optimal Energy Efficient Design of Artificial Noise
  to Prevent Side-Channel Attacks}. In \bibinfo{booktitle}{\emph{2022 IEEE
  International Workshop on Information Forensics and Security~(WIFS'22)}}.
  \bibinfo{pages}{1--6}.
\newblock


\bibitem[Kocher et~al\mbox{.}(1999)]%
        {kocher1999differential}
\bibfield{author}{\bibinfo{person}{Paul Kocher}, \bibinfo{person}{Joshua
  Jaffe}, {and} \bibinfo{person}{Benjamin Jun}.}
  \bibinfo{year}{1999}\natexlab{}.
\newblock \showarticletitle{{Differential Power Analysis}}. In
  \bibinfo{booktitle}{\emph{Advances in cryptology (CRYPTO'99)}}. Springer,
  \bibinfo{publisher}{Springer}, \bibinfo{pages}{789--789}.
\newblock


\bibitem[Levey and McLaughlin(2002)]%
        {levery2002statistical}
\bibfield{author}{\bibinfo{person}{David~B. Levey} {and}
  \bibinfo{person}{Stephen McLaughlin}.} \bibinfo{year}{2002}\natexlab{}.
\newblock \showarticletitle{{The Statistical Nature of Impulse Noise
  Interarrival Times in Digital Subscriber Loop Systems}}.
\newblock \bibinfo{journal}{\emph{Signal Processing}} \bibinfo{volume}{82},
  \bibinfo{number}{3} (\bibinfo{year}{2002}), \bibinfo{pages}{329--351}.
\newblock


\bibitem[Mangard et~al\mbox{.}(2008)]%
        {mangard2008power}
\bibfield{author}{\bibinfo{person}{Stefan Mangard}, \bibinfo{person}{Elisabeth
  Oswald}, {and} \bibinfo{person}{Thomas Popp}.}
  \bibinfo{year}{2008}\natexlab{}.
\newblock \bibinfo{booktitle}{\emph{{Power Analysis Attacks: Revealing the
  Secrets of Smart Cards}}}.
\newblock \bibinfo{publisher}{Springer}.
\newblock


\bibitem[Mann et~al\mbox{.}(2002)]%
        {mann2002impulse}
\bibfield{author}{\bibinfo{person}{Iain Mann}, \bibinfo{person}{Stephen
  McLaughlin}, \bibinfo{person}{Werner Henkel}, \bibinfo{person}{Rob Kirkby},
  {and} \bibinfo{person}{Thomas Kessler}.} \bibinfo{year}{2002}\natexlab{}.
\newblock \showarticletitle{{Impulse Generation With Appropriate Amplitude,
  Length, Inter-arrival, and Spectral Characteristics}}.
\newblock \bibinfo{journal}{\emph{IEEE Journal on Selected Areas in
  Communications}} \bibinfo{volume}{20}, \bibinfo{number}{5}
  (\bibinfo{year}{2002}), \bibinfo{pages}{901--912}.
\newblock


\bibitem[Nemhauser and Wolsey(1988)]%
        {Nemhauser1988integer}
\bibfield{author}{\bibinfo{person}{George~L. Nemhauser} {and}
  \bibinfo{person}{Laurence~A. Wolsey}.} \bibinfo{year}{1988}\natexlab{}.
\newblock \bibinfo{booktitle}{\emph{{Integer and Combinatorial Optimization}}}.
\newblock \bibinfo{publisher}{Wiley-Interscience}.
\newblock
\showISBNx{047182819X}


\bibitem[Prouff and Rivain(2013)]%
        {prouff2013masking}
\bibfield{author}{\bibinfo{person}{Emmanuel Prouff} {and}
  \bibinfo{person}{Matthieu Rivain}.} \bibinfo{year}{2013}\natexlab{}.
\newblock \showarticletitle{{Masking against Side-Channel Attacks: A Formal
  Security Proof}}. In \bibinfo{booktitle}{\emph{32nd Annual International
  Conference on the Theory and Applications of Cryptographic
  Techniques~(Eurocrypt'13)}}. \bibinfo{publisher}{Springer},
  \bibinfo{pages}{142--159}.
\newblock


\bibitem[Schindler et~al\mbox{.}(2005)]%
        {schindler2005stochastic}
\bibfield{author}{\bibinfo{person}{Werner Schindler}, \bibinfo{person}{Kerstin
  Lemke}, {and} \bibinfo{person}{Christof Paar}.}
  \bibinfo{year}{2005}\natexlab{}.
\newblock \showarticletitle{{A Stochastic Model for Differential Side Channel
  Cryptanalysis}}. In \bibinfo{booktitle}{\emph{International Workshop on
  Cryptographic Hardware and Embedded Systems~(CHES'05)}}. Springer,
  \bibinfo{pages}{30--46}.
\newblock


\bibitem[Shafiee and Ulukus(2005)]%
        {sha2005jam}
\bibfield{author}{\bibinfo{person}{Shabnam Shafiee} {and}
  \bibinfo{person}{Sennur Ulukus}.} \bibinfo{year}{2005}\natexlab{}.
\newblock \showarticletitle{{Capacity of Multiple Access Channels with
  Correlated Jamming}}. In \bibinfo{booktitle}{\emph{2005 IEEE Military
  Communications Conference'(MILCOM'05)}}, Vol.~\bibinfo{volume}{1}.
  \bibinfo{publisher}{IEEE}, \bibinfo{pages}{218--224}.
\newblock


\bibitem[Shamir(2000)]%
        {protect2000shamir}
\bibfield{author}{\bibinfo{person}{Adi Shamir}.}
  \bibinfo{year}{2000}\natexlab{}.
\newblock \showarticletitle{P{rotecting Smart Cards from Passive Power Analysis
  with Detached Power Supplies}}. In \bibinfo{booktitle}{\emph{International
  Workshop on Cryptographic Hardware and Embedded Systems~(CHES'00)}}.
  \bibinfo{publisher}{Springer}, \bibinfo{pages}{1--77}.
\newblock


\bibitem[Standaert and Archambeau(2008)]%
        {standaert2008using}
\bibfield{author}{\bibinfo{person}{Fran{\c{c}}ois-Xavier Standaert} {and}
  \bibinfo{person}{C{\'e}dric Archambeau}.} \bibinfo{year}{2008}\natexlab{}.
\newblock \showarticletitle{{Using Subspace-based Template Attacks to Compare
  and Combine Power and Electromagnetic Information Leakages}}. In
  \bibinfo{booktitle}{\emph{International Workshop on Cryptographic Hardware
  and Embedded Systems~(CHES'08)}}. \bibinfo{publisher}{Springer},
  \bibinfo{pages}{411--425}.
\newblock


\bibitem[Standaert et~al\mbox{.}(2009)]%
        {standaert2009unified}
\bibfield{author}{\bibinfo{person}{Fran{\c{c}}ois-Xavier Standaert},
  \bibinfo{person}{Tal Malkin}, {and} \bibinfo{person}{Moti Yung}.}
  \bibinfo{year}{2009}\natexlab{}.
\newblock \showarticletitle{{A Unified Framework for the Analysis of
  Side-Channel Key Recovery Attacks}}. In \bibinfo{booktitle}{\emph{28th Annual
  International Conference on the Theory and Applications of Cryptographic
  Techniques~(Eurocrypt'09)}}. \bibinfo{publisher}{Springer},
  \bibinfo{pages}{443--461}.
\newblock


\bibitem[Unterluggauer et~al\mbox{.}(2018)]%
        {leakage2018thomas}
\bibfield{author}{\bibinfo{person}{Thomas Unterluggauer},
  \bibinfo{person}{Thomas Korak}, \bibinfo{person}{Stefan Mangard},
  \bibinfo{person}{Robert Schilling}, \bibinfo{person}{Luca Benini},
  \bibinfo{person}{Frank~K. G{\"u}rkaynak}, {and} \bibinfo{person}{Michael
  Muehlberghuber}.} \bibinfo{year}{2018}\natexlab{}.
\newblock \showarticletitle{{Leakage Bounds for Gaussian Side Channels}}. In
  \bibinfo{booktitle}{\emph{International Conference on Smart Card Research and
  Advanced Applications~(CARDIS'18)}}. \bibinfo{publisher}{Springer},
  \bibinfo{pages}{88--104}.
\newblock


\bibitem[Whitnall and Oswald(2015)]%
        {whitnall2015robust}
\bibfield{author}{\bibinfo{person}{Carolyn Whitnall} {and}
  \bibinfo{person}{Elisabeth Oswald}.} \bibinfo{year}{2015}\natexlab{}.
\newblock \showarticletitle{{Robust Profiling for DPA-Style Attacks}}. In
  \bibinfo{booktitle}{\emph{International Workshop on Cryptographic Hardware
  and Embedded Systems~(CHES'15)}}. \bibinfo{publisher}{Springer},
  \bibinfo{pages}{3--21}.
\newblock


\bibitem[Zhou et~al\mbox{.}(2017)]%
        {Zhou2017ANU}
\bibfield{author}{\bibinfo{person}{Xinping Zhou}, \bibinfo{person}{Carolyn
  Whitnall}, \bibinfo{person}{Elisabeth Oswald}, \bibinfo{person}{Degang Sun},
  {and} \bibinfo{person}{Zhu Wang}.} \bibinfo{year}{2017}\natexlab{}.
\newblock \showarticletitle{{A Novel Use of Kernel Discriminant Analysis as a
  Higher-Order Side-Channel Distinguisher}}. In
  \bibinfo{booktitle}{\emph{International Conference on Smart Card Research and
  Advanced Applications~(CARDIS'17)}}. \bibinfo{publisher}{Springer},
  \bibinfo{pages}{70--87}.
\newblock


\end{thebibliography}

\end{document}